\documentclass[prd,aps,floats,floatfix,nofootinbib,preprintnumbers]{revtex4-1}
\usepackage{amsmath}
\usepackage{amsfonts}
\usepackage{graphicx}
\usepackage{slashed}
\usepackage{dcolumn}
\usepackage{bm}
\usepackage{amssymb}
\usepackage{latexsym}
\usepackage{color}

\newcommand{\brac}[1]{\left( #1 \right)}

\begin{document}

\title{Hadron Collider Production of Massive Color-Octet Vector Bosons at Next-to-Leading Order}

\author{R. Sekhar Chivukula}
\email[]{sekhar@msu.edu}
\affiliation{Department of Physics and Astronomy, Michigan State University, East Lansing, MI 48824, USA}
\author{Arsham Farzinnia}
\email[]{farzinnia@tsinghua.edu.cn}
\affiliation{Institute of Modern Physics, Center for High Energy Physics Research, Tsinghua University, Beijing 100084, China}
\author{Jing Ren}
\email[]{renj08@mails.tsinghua.edu.cn}
\affiliation{Institute of Modern Physics, Center for High Energy Physics Research, Tsinghua University, Beijing 100084, China}
\author{Elizabeth H. Simmons}
\email[]{esimmons@msu.edu}
\affiliation{Department of Physics and Astronomy, Michigan State University, East Lansing, MI 48824, USA}

\date{\today}

\begin{abstract}
This paper completes the study of the next-to-leading order (NLO) QCD corrections to massive color-octet vector boson production at the LHC and Tevatron. The massive color-octet vector bosons are generically referred to as colorons. Building on our previous calculation of quark-initiated coloron production at NLO, we use the pinch technique to investigate coloron production via gluon fusion. We demonstrate that this one-loop production amplitude is finite, and find that its numerical contribution to coloron production is typically four orders of magnitude smaller than the contribution from quark annihilation. Coloron production via gluon fusion is therefore only relevant if the colorons are (nearly) fermiophobic. We then present extensive plots and tables of our full results for NLO coloron production at the Tevatron and the LHC.
\end{abstract}

\maketitle

\section{Introduction}\label{intro}

This paper completes the study of the next-to-leading order (NLO) QCD corrections to 
the production of massive vector color-octet bosons at the LHC and Tevatron begun in \cite{Chivukula:2011ng}. 
In this analysis, we use the generic name ``colorons'' to refer to massive color-octet vector bosons, regardless of the details of their couplings to fermions.
Colorons are a feature of a variety of models, 
including axigluon models \cite{Frampton:1987dn,Bagger:1987fz}, topcolor models \cite{Hill:1991at,Hill:1993hs,Popovic:1998vb,Braam:2007pm}, technicolor models with colored technifermions \cite{Chivukula:1995dt}, flavor-universal \cite{Chivukula:1996yr,Simmons:1996fz} and chiral \cite{Martynov:2009en} coloron models, and extra-dimensional models with KK gluons \cite{Davoudiasl:2000wi,Lillie:2007yh}. 
The tree-level hadron-collider phenomenology of colorons has been discussed extensively in various experimental and theoretical
contexts, see \cite{Dicus:1994sw,Dobrescu:2007yp,Kilic:2008pm,Kilic:2008ub,Han:2010rf,Dicus:2010bm,Sayre:2011ed,Bai:2011mr}
and references therein.
Searches for resonances in the dijet mass spectrum at the LHC are now, depending on the details of
how the coloron couples, sensitive up to coloron masses of order  4-5 TeV \cite{Chatrchyan:2013qha,ATLAS:2012pu}.\footnote{At least for the fermion charge assignments considered, and in the case where the resonance is narrow compared to the djiet mass resolution of the detector}  If there are color-octet vector bosons associated with the electroweak symmetry breaking sector, as suggested by several of  the models mentioned above, their presence should be uncovered by the LHC.

In particular, in this paper we provide a complete and consistent calculation of the gluon fusion contribution to coloron production. This amplitude vanishes at tree level \cite{Chivukula:2001gv}, and therefore occurs initially at one-loop and is finite.
Despite the large gluon parton luminosity at the Tevatron and the LHC, we show that the gluon-fusion contribution to coloron production is typically four orders of magnitude smaller than the contribution from quark annihilation.\footnote{In the special case of the axigluon \protect\cite{Frampton:1987dn,Bagger:1987fz}, the gluon fusion contribution to coloron production is forbidden by a discrete symmetry of the theory. See Sec. \protect\ref{GGCHadron}.}   Coloron production via gluon fusion is therefore only relevant if the colorons are (nearly) fermiophobic, with quark couplings
of order $10^{-2} g_s$ or smaller. 

We also update previous results on the NLO K-factor for coloron production using more modern structure functions (CT10 \cite{Lai:2010vv}) and present extensive plots and tables of our results for NLO coloron production at the Tevatron and at the LHC for energies of  $\sqrt{s}=7$, 8, and 14 TeV.

Following \cite{Chivukula:2011ng}, we compute the gauge-, quark-, and self-couplings of the coloron from a theory with
an extended $SU(3)_{1c} \times SU(3)_{2c} \to SU(3)_c$ gauge structure, where $SU(3)_c$ is identified with QCD.
As noted there, the self-couplings of KK gluons in extra-dimensional models, or of colored technivector mesons in
technicolor models, will not precisely follow this pattern. Our calculations continue to apply approximately to these cases, however, 
to the extent that the $SU(3)_{1c} \times SU(3)_{2c}$
model is a good low-energy effective theory for the extra dimensional model (a ``two-site" approximation in
the language of deconstruction \cite{ArkaniHamed:2001ca,Hill:2000mu}) or for the technicolor theory (a hidden local symmetry approximation for
the effective technivector meson sector \cite{Bando:1984ej,Bando:1985rf}).

Our calculation of the gluon fusion contribution to coloron production uses the ``pinch" technique \cite{Binosi:2009qm} to simplify the calculation and demonstrate the gauge-invariance of our result. Previous considerations of this process 
were either incomplete \cite{Djouadi:2007eg}, ignoring the contribution from vector-boson loops, or assumed color gauge-invariance without explicitly checking the consistency of the calculation \cite{Allanach:2009vz}. While our results confirm that the contribution of gluon fusion to coloron production is small, we find that the contribution is typically an order of magnitude smaller than reported in \cite{Allanach:2009vz}.

The paper is structured as follows. In Sec. \ref{model} we describe the $SU(3)_{1c} \times SU(3)_{2c}$ model and establish our notation. Our conventions follow those introduced in \cite{Chivukula:2011ng} and, for convenience, the Feynman rules for the model are noted in Appendix \ref{FR}. In Sec. \ref{GGC} we describe the calculation of the one-loop partonic amplitude for the gluon fusion contribution to coloron production in detail. In Sec. \ref{GGCHadron} we compute the size of the gluon fusion contribution to coloron production at the Tevatron and LHC. In Sec. \ref{Kfactors} we provide the update of the NLO K-factors for coloron production originally reported in \cite{Chivukula:2011ng}, extending that work to the Tevatron and to the LHC at $\sqrt{s}=7$, 8, and 14 TeV. Tables of the numerical values of the K-factors are provided in Appendix \ref{Kf}. Finally, Sec. \ref{conclude} presents our conclusions.

\section{Formalism}\label{model}

We consider color-octet vector bosons, which arise from an extended color gauge group $SU(3)_{1c} \times SU(3)_{2c}$, spontaneously broken to the diagonal subgroup, $SU(3)_{c}$.\footnote{Throughout this paper, we closely follow the formalism and notation introduced in \cite{Chivukula:2011ng}.} The latter is identified with the ordinary QCD. The symmetry breaking produces, in addition to the massless color-octet of gluons, a massive color-octet of vector bosons, which we generically refer to as {\em colorons}. We model the symmetry breaking sector minimally, using a non-linear sigma model.\footnote{The additional particles present in a more complete theory, e.g. additional scalars arising from a linear sigma model, will not fundamentally change the calculations presented here.}

The Lagrangian of the $SU(3)_{1c} \times SU(3)_{2c}$ model, with the couplings $g_{s_{1}}$ and $g_{s_{2}}$ respectively, is given by
\begin{equation} \label{Lagr}
\mathcal{L}_{\text{color}} = - \frac{1}{2} \text{Tr} \left [ G_{1 \mu \nu} G_{1}^{ \mu \nu} \right ] - \frac{1}{2} \text{Tr} \left [ G_{2 \mu \nu} G_{2}^{ \mu \nu} \right ] +\frac{f^2}{4}\ {\rm Tr} D_\mu\Sigma \,D^\mu\Sigma^\dagger + {\cal L}_{\rm gauge-fixing} + {\cal L}_{\rm ghost} + {\cal L}_{\rm quark} \ .
\end{equation}
Here, $\Sigma$ is the non-linear sigma field inducing the symmetry breaking,
\begin{equation} \label{sigma}
\Sigma = \exp\left(\frac{2 i \pi^a t^a}{f}\right) \ , \quad a=1,\dots,8 \ ,
\end{equation}
where $\pi^a$ are the Nambu-Goldstone bosons `eaten' by the colorons, and $f$ is the corresponding `decay-constant'. The $\Sigma$ field transforms as the bi-fundamental of $SU(3)_{1c}\times SU(3)_{2c}$,
\begin{equation}
\Sigma\to u_1 \Sigma u_2^\dagger \ , \quad u_i=\exp\left(i \alpha_i^a t^a\right) \ ,
\end{equation}
where the $\alpha_i^a$ are the parameters of the original $SU(3)_{ic}$ transformations. We have then the covariant derivative
\begin{equation}\label{covder}
D_\mu \Sigma = \partial_\mu \Sigma - i g_{s_1} G^a_{1\mu} t^a \Sigma + i g_{s_2} \Sigma\,  G^a_{2\mu} t^a \ , \qquad t^{a} = \frac{\lambda^{a}}{2} \ ,
\end{equation}
with $\lambda^{a}$ the Gell-Mann matrices, and ${\rm Tr}\ t^a t^b=\delta^{ab}/2$. 

Using the procedure outlined in \cite{Chivukula:2011ng}, one may diagonalize the mass term in the quadratic Lagrangian by means of an orthogonal rotation, $S$.\footnote{The ${\cal L}_{\rm gauge-fixing}$ and ${\cal L}_{\rm ghost} $ terms are also constructed in \cite{Chivukula:2011ng}. We merely employ the results found in there, and ultimately perform the calculations in the 't Hooft-Feynman gauge, $\xi = 1$.} This leads to the following definitions in the mass eigenstate basis
\begin{equation}\label{eq:rotations}
\left(\begin{array}{c} G^a_{1\mu} \\ G^a_{2\mu} \end{array}\right) = S \left(\begin{array}{c} G^a_\mu \\ C^a_\mu \end{array}\right) \ ,
\end{equation}
with the mixing defined by the angle $\theta_c$ and 
\begin{equation}
S\equiv \left(\begin{array}{cc}\cos\theta_c & -\sin\theta_c \\ \sin\theta_c & \cos\theta_c \end{array}\right) \ , \quad
\sin\theta_c \equiv \frac{g_{s_1}}{\sqrt{g_{s_1}^2+g_{s_2}^2}} \ .
\end{equation}
In \eqref{eq:rotations}, $G^a_\mu$ is defined as the gluon field and $C^a_\mu$ represents the coloron. The gluon is massless, whereas the coloron acquires a mass
\begin{equation}
M_C = \frac{\sqrt{g_{s_1}^2+g_{s_2}^2}\, f}{2}\equiv \frac{g_s\, f}{\sin 2\theta_c} \ ,
\end{equation}
with $g_s$ the QCD $SU(3)_c$ coupling,
\begin{equation}
\frac{1}{g_s^2} = \frac{1}{g_{s_1}^2}+\frac{1}{g_{s_2}^2}\ .
\end{equation}

In order to keep the current description applicable to a wide variety of the available models, we assign arbitrary matter couplings to the extended gauge group, $SU(3)_{1c} \times SU(3)_{2c}$, in accordance with the treatment in \cite{Chivukula:2011ng}. As a consequence, in the mass eigenstate basis, we may write
\begin{equation} \label{Lferm}
\mathcal{L}_{\rm quark} = \bar{q}^k i\left[\slashed{\partial}-i g_s \slashed{G}^a t^a -i \slashed{C}^a t^a \left(g_L P_L+g_R P_R\right)\right]q_k \ ,
\end{equation}
where $P_L$ and $P_R$ are the helicity projection operators,
\begin{equation}\label{PLPR}
P_{L,R}\equiv \frac{1\mp \gamma_5}{2}\ ,
\end{equation}
and $k$ is a flavor index (implicit summation assumed).\footnote{Here we work in the broken electroweak phase, and only employ fermion mass eigenstates.} The coupling of the quarks to the gluon is, as usual, dictated by charge universality. The quark couplings to the coloron, on the other hand, may be chosen to be chiral, and is in general parametrized by the $g_L$ and $g_R$ couplings, representing respectively the couplings of the left- and right-handed quarks to the coloron. These couplings depend on how one decides to charge the quarks under the original extended gauge group. In general, each of the $g_L$ and $g_R$ parameters in any specific model can take on the values\footnote{Generally speaking, both $g_L$ and $g_R$ should be matrices in flavor space. For simplicity, in what follows, we assume that the coloron couplings are flavor-universal -- the generalization to the flavor-dependent case is straightforward. Flavor-changing couplings are, however, strongly constrained \cite{Chivukula:2010fk}.}
\begin{equation} \label{gLgR}
g_L, g_R \in \left \{ -g_s \tan \theta_c, g_s \cot \theta_c \right \} \ .
\end{equation}
For example, if both left-handed and right-handed quarks are only charged under $SU(3)_{2c}$, then $g_L=g_R=g_s\cot\theta_c$, resulting in a vector-like theory. A chiral example would be the axigluon (with the maximal mixing, $\theta_c=\pi/4$) \cite{Frampton:1987dn,Bagger:1987fz}, which corresponds to $g_L=-g_R=g_s$. For later convenience, we introduce the coefficients
\begin{equation}\label{rLrR}
r_L\equiv \frac{g_L}{g_s}\ , \quad r_R\equiv \frac{g_R}{g_s} \ , \qquad r_L, r_R \in \left \{-\tan \theta_c, \cot \theta_c \right \} \ .
\end{equation}
The interaction vertices and the corresponding Feynman rules are presented in the Appendix~\ref{FR}.

\section{Coloron Production Via Gluon Fusion}\label{GGC}

Gluon-initiated production of colorons cannot proceed at tree-level, as there are no gauge-invariant dimension-four operators accommodating such interaction (see the Feynman rules in the Appendix~\ref{FR}): in general, there are no dimension-four terms with two gauge bosons of an unbroken symmetry and a vector field charged under the same symmetry.\footnote{The lowest available operators accounting for such a process are of dimension six \cite{Chivukula:2001gv}.} At one-loop, however, a coloron can be produced by fusing two initial-state gluons. Naively, one might expect the cross section of this process to be negligible as compared with the tree-level quark-initiated production channel and its higher-order contributions. However, at a high-CM energy hadron collider, such as the LHC, there are many more gluons available to facilitate this fusion process. It is, therefore, important to investigate the rate of this production channel, and draw quantitative comparisons with the tree-level induced process which proceeds via the quark-antiquark pair annihilation.

In this section, we compute the leading order amplitude for coloron production via gluon fusion, which is induced at one-loop. We employ the narrow-width approximation for the coloron, take all the external gauge bosons to be on-shell, and set the quark masses to zero: given the current TeV lower bounds on the coloron mass 
\cite{Chatrchyan:2013qha,ATLAS:2012pu}
these approximations are justified even for the top-quark.

\begin{figure}
\begin{center}
\includegraphics[width=.25\textwidth]{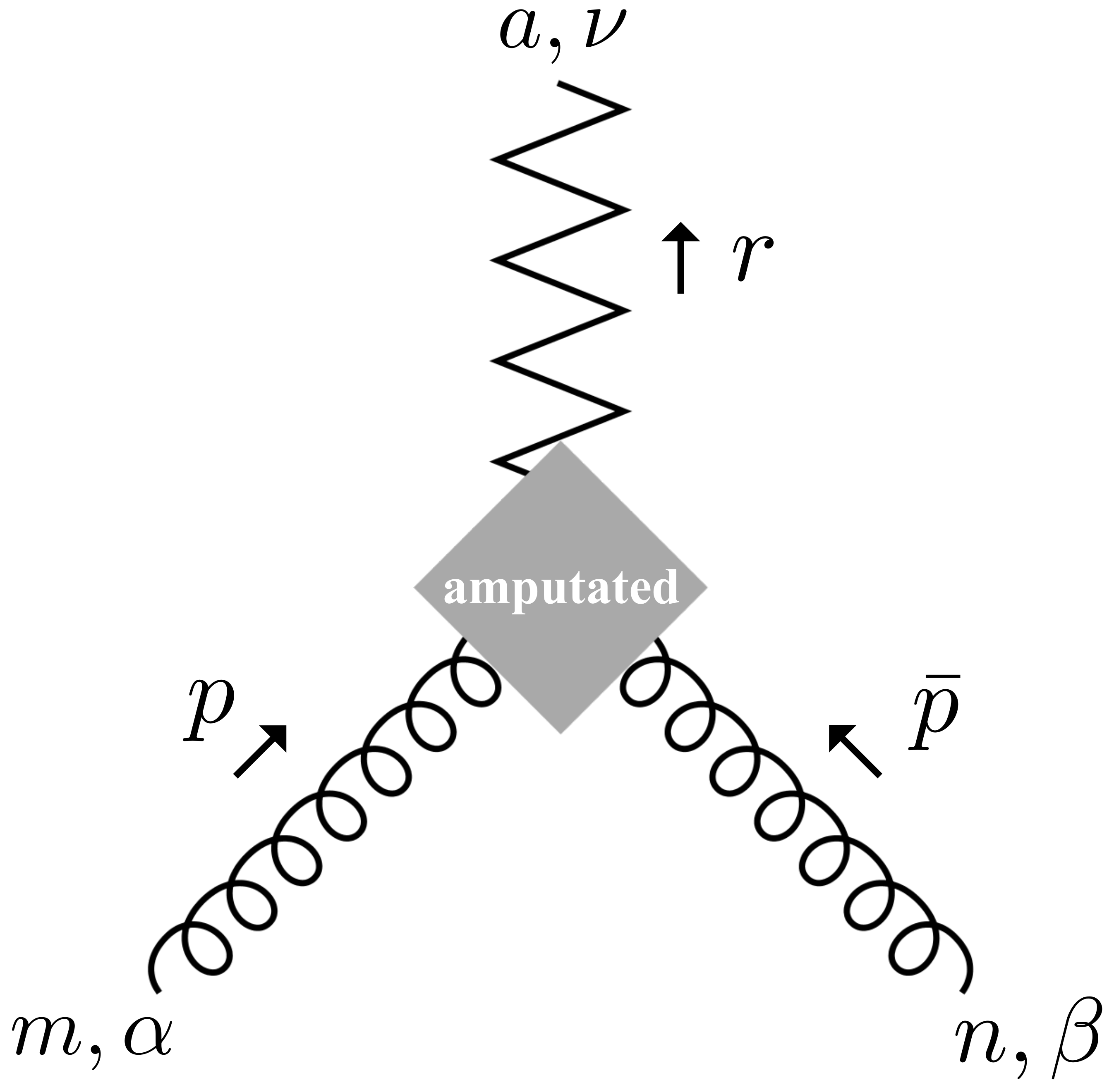}
\caption{The vertex contributions to the gluon-initiated coloron production. The diamond represents the amputated vertex diagrams, and the coloron is illustrated by the zigzag line. The explicit kinematics for the external legs with the corresponding Lorentz (Greek) and color (Latin) indices are displayed.}
\label{LSZ}
\end{center}
\end{figure}

\begin{figure}
\begin{center}
\includegraphics[width=.8\textwidth]{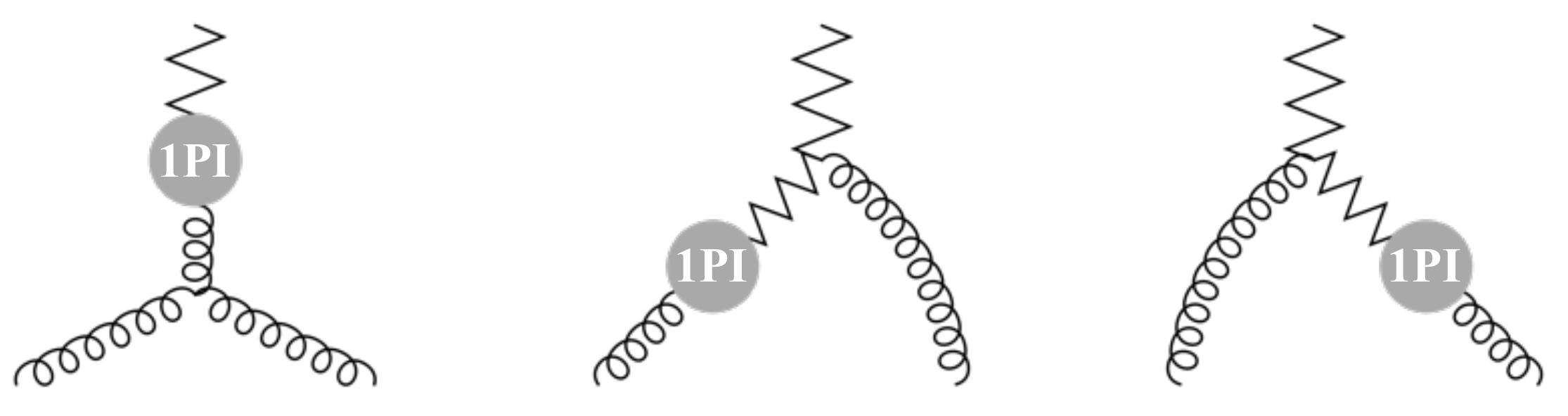}
\caption{Gluon-fusion to coloron via the one-particle irreducible mixed vacuum polarization amplitudes (VPA).}
\label{gaugeS}
\end{center}
\end{figure}

Let us define the amplitude of this process (see Fig.~\ref{LSZ} for the explicit kinematics) as
\begin{equation}\label{iM}
i\mathcal{M}_{gg\to C} \equiv (iS + iV + iS_{\text{ferm}} + iV_{\text{ferm}}) \times \varepsilon^{* a \, (\lambda_{C})}_{C \, \nu} (r) \, \varepsilon^{m \, (\lambda_{g_1})}_{g \, \alpha} (p) \, \varepsilon^{n \, (\lambda_{g_2})}_{g \, \beta} (\bar{p}) \ ,
\end{equation}
where $\lambda_{C(g)}$ is the coloron (gluon) polarization, and Greek (Latin) letters denote Lorentz (color) indices. $iV$ and $iV_{\text{ferm}}$ are the amputated one-loop vertices (Fig.~\ref{LSZ}) originating from the gauge and matter sectors of the theory, respectively, whereas $iS$ and $iS_{\text{ferm}}$ form the corresponding contributions from the (mixed) vacuum polarization amplitude diagrams (Fig.~\ref{gaugeS}). The calculations are performed in 't Hooft-Feynman gauge, $\xi=1$, using dimensional regularization, and, as we shall demonstrate, construct a UV- and IR-finite one-loop amplitude. In the final result, we employ the on-shell identities
\begin{equation}\label{EOM}
\begin{split}
& r^2 = M_C^2 \ , \quad r \cdot \varepsilon_{C} (r) = (p+\bar p) \cdot \varepsilon_{C} (r) = 0 \ , \\
& p^2=\bar p^2 = 0 \ , \quad p \cdot \varepsilon_{g} (p) = \bar{p} \cdot \varepsilon_{g} (\bar{p}) = 0 \ .
\end{split}
\end{equation}

\begin{figure}
\begin{center}
\includegraphics[width=.25\textwidth]{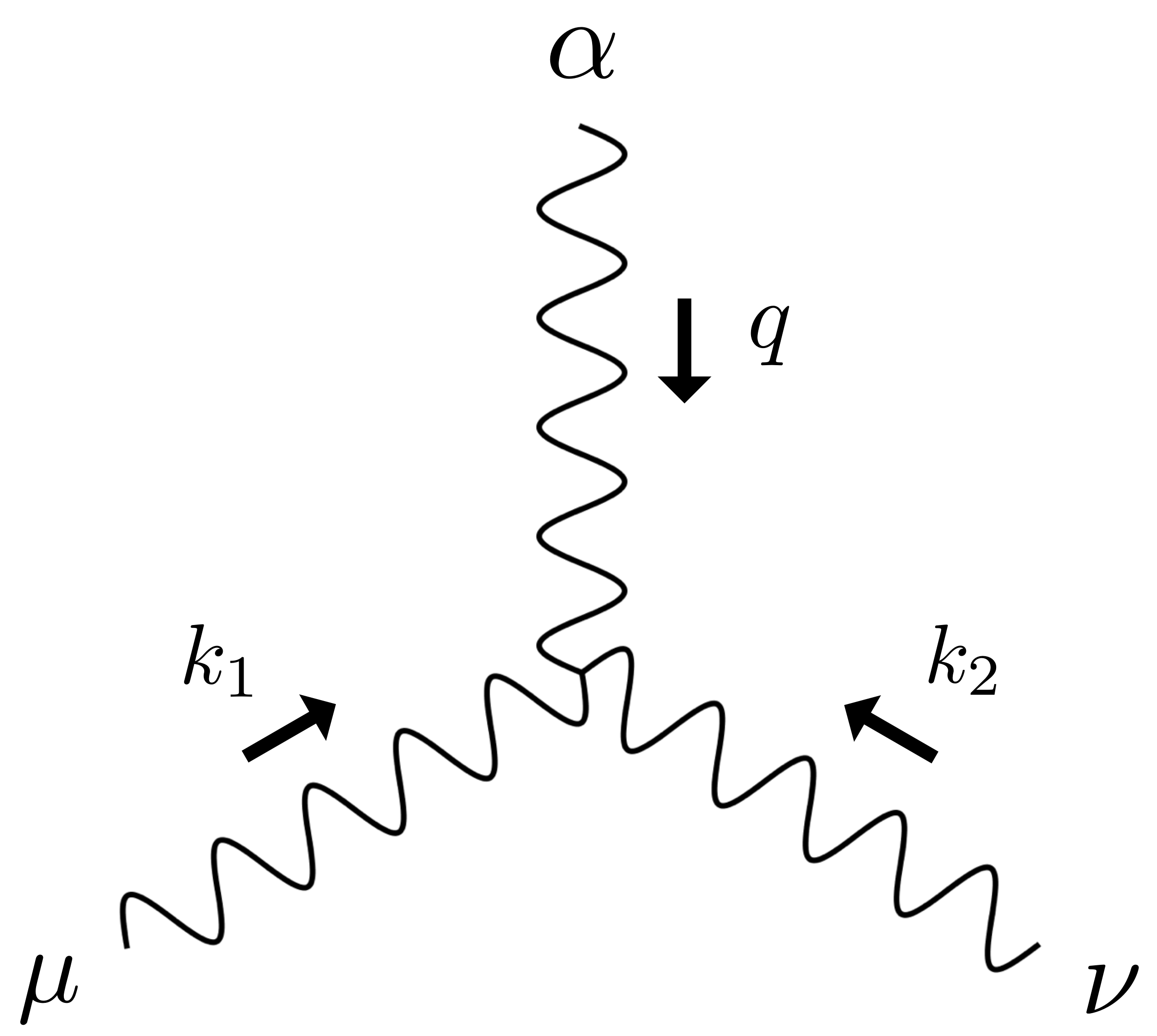}
\caption{Kinematics of a generic triple-gauge boson vertex, with the Lorentz structure given by $\Gamma_{\alpha\mu\nu}(q,k_1,k_2)$ \eqref{Gamma}. Each leg could be either a gluon or a coloron, depending on the specific vertex (see the Feynman rules in Appendix~\ref{FR}). All momenta flow towards the vertex.}
\label{3G}
\end{center}
\end{figure}

Furthermore, we utilize the \textit{Pinch Technique} \cite{Binosi:2009qm} to demonstrate that the full gluon-fusion to coloron amplitude satisfies QED-like Ward identities, and is therefore gauge invariant and finite at one-loop. In the context of our work, the pinch technique amounts to decomposing an arbitrary non-Abelian triple-gauge boson vertex (Fig.~\ref{3G}) with the Lorentz structure
\begin{equation}\label{Gamma}
\Gamma_{\alpha\mu\nu}(q,k_1,k_2) \equiv g_{\mu\nu}(k_1-k_2)_\alpha + g_{\alpha\nu}(k_2-q)_\mu + g_{\alpha\mu}(q-k_1)_\nu \ ,
\end{equation}
into two parts
\begin{equation}\label{dec}
\Gamma_{\alpha\mu\nu}(q,k_1,k_2) = \Gamma^F_{\alpha\mu\nu}(q,k_1,k_2) + \Gamma^P_{\alpha\mu\nu}(q,k_1,k_2) \ ,
\end{equation}
where
\begin{equation}\label{FandP}
\begin{split}
\Gamma^F_{\alpha\mu\nu}(q,k_1,k_2) \equiv&\, (k_1-k_2)_\alpha g_{\mu\nu} + 2q_\nu g_{\alpha\mu} - 2q_\mu g_{\alpha\nu} \ , \\
\Gamma^P_{\alpha\mu\nu}(q,k_1,k_2) \equiv&\, k_{2\nu} g_{\alpha\mu} - k_{1\mu} g_{\alpha\nu} \ .
\end{split}
\end{equation}
Unlike $\Gamma_{\alpha\mu\nu}(q,k_1,k_2)$, the $\Gamma_{\alpha\mu\nu}^F(q,k_1,k_2)$ vertex satisfies a QED-like Ward identity
\begin{equation}\label{WI}
q^\alpha \Gamma^F_{\alpha\mu\nu}(q,k_1,k_2) = (k_2^2-k_1^2) g_{\mu\nu} \ .
\end{equation}

In what follows, we study the calculation of one-loop diagrams containing gauge boson trilinear vertices such as \eqref{Gamma}. We will illustrate how replacing $\Gamma_{\alpha\mu\nu}(q,k_1,k_2)$ by $\Gamma_{\alpha\mu\nu}^F(q,k_1,k_2)$ corresponds to identifying `unpinched' Feynman diagrams. The virtue lies in the observation that these diagrams satisfy QED-like Ward identities, {\em a la} \eqref{WI}, and are kept as part of the original calculation. The remaining piece of the vertex, $\Gamma_{\alpha\mu\nu}^P(q,k_1,k_2)$, will remove an internal propagator from the original diagram, giving rise to `pinched' Feynman diagrams. These pinched structures, possessing a different topology than their primitive diagrams, may then be removed from the original calculation and reassigned to the computation of diagrams elsewhere in the theory that share the same topology. This leaves behind only the gauge-invariant unpinched contribution in the original calculation. As such, employing the pinch technique allows one to exhibit the consistent renormalization and gauge-invariance of the one-loop amplitude in a non-Abelian theory, by recovering QED-like Ward identities through a systematic reshuffling of different terms within the amplitude.\footnote{See \cite{Binosi:2009qm} and references therein for a thorough review of the pinch technique and its applications.} Throughout the paper, we denote the unpinched and pinched diagrams symbolically by inserting a black disk over their relevant vertices, in order to distinguish them from their primitive diagrams.

\subsection{Vacuum Polarization Amplitudes {\em a la} Pinch Technique: Gauge Sector}\label{gVPA}

The gluon fusion to coloron process facilitated by the vacuum polarization amplitudes that mix colorons with gluons (henceforth called mixed-VPA) is shown in Fig.~\ref{gaugeS}. It is of the form
\begin{equation}\label{iS}
iS = g_s \left\{ f^{dmn} \Gamma^{\rho\alpha\beta}(-r,p,\bar{p}) \, \frac{\Pi^{ad}_{\nu\rho} (r)}{r^2} + f^{adn} \Gamma^{\nu\rho\beta}(-r,p,\bar{p}) \, \frac{\Pi^{md}_{\alpha\rho} (p)}{p^2-M_C^2} + f^{amd} \Gamma^{\nu\alpha\rho}(-r,p,\bar{p}) \,\frac{\Pi^{nd}_{\beta\rho} (\bar{p})}{\bar{p}^2-M_C^2} \right\} \ ,
\end{equation}
with the definition \eqref{Gamma} assumed. Notice the presence of the internal gauge boson propagators in this expression: $1/r^2$~associated with an internal gluon exchange, $1/ (p^2 - M_C^2)$ and $1/(\bar p^2 - M_C^2)$ associated with an internal coloron exchange. The symbol $f^{abc}$~represents the $SU(3)_c$ antisymmetric structure constant.

\begin{figure}
\begin{center}
\includegraphics[width=.95\textwidth]{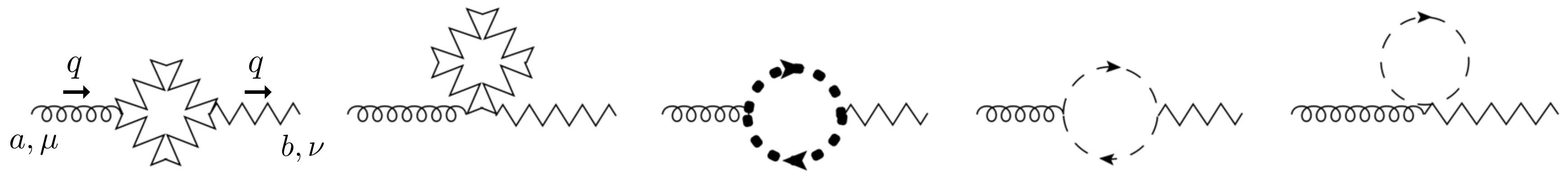}
\caption{The gauge sector 1PI gluon-coloron mixing at one-loop. The curly lines represent gluons, the zig-zag lines represent colorons, the heavy dots are the coloron ghosts, while the eaten Nambu-Goldstone boson is represented by dashed lines.}
\label{mVPA}
\end{center}
\end{figure}

The $i\Pi^{ab}_{\mu\nu} (q)$ term denotes the sum of the gauge sector one-particle irreducible (1PI) mixed-VPA diagrams with external momentum $q$ (Fig.~\ref{mVPA}); in $d=4-2\epsilon$ dimensions it reads
\begin{equation}\label{iPraw}
\begin{split}
i\Pi^{ab}_{\mu\nu} (q) = g_s^2 \cot (2\theta_c) \, C_2(G) \delta^{ab}\, \mu^{4-d}\int \frac{d^d k}{(2\pi)^d} \left\{ \frac{\Gamma_1\Gamma_2-2k_\mu(k+q)_\nu+\frac{1}{2}(2k+q)_\mu(2k+q)_\nu}{\left[k^2-M_C^2+i\eta \right]\left[(k+q)^2-M_C^2+i\eta \right]} + \frac{(1-2d) g_{\mu\nu}}{k^2-M_C^2+i\eta} \right\} \ ,
\end{split} 
\end{equation}
where the parameter $\mu$ is the mass scale introduced by the loop integral in $d$ dimensions, $\eta$ is the positive infinitesimal parameter giving the appropriate prescription for computing the integral in momentum space, and $C_2(G)=3$ is the Casimir of the adjoint representation. Here, $k$ represents the momentum running in the loop, and we have employed the color product identity
\begin{equation}\label{ffid}
f^{acd}f^{bcd} =  C_2(G) \delta^{ab} \ .
\end{equation}
In addition, we have defined the two non-Abelian vertices contained in the first diagram of Fig.~\ref{mVPA} as $\Gamma_1$ and $\Gamma_2$ (suppressing the Lorentz indices),
\begin{equation}\label{VPAgam}
\Gamma_1 \equiv \Gamma^{\;\lambda \rho}_\nu(q,-(k+q),k) \ , \qquad \Gamma_2 \equiv \Gamma_{\mu\lambda \rho}(q,-(k+q),k) \ ,
\end{equation}
with again the definition \eqref{Gamma} implied.

A straightforward calculation of \eqref{iPraw}, however, reveals an overall non-transverse structure for the mixed-VPA, which might raise concerns since it obscures the gauge invariance of the theory.\footnote{To be specific, the overall form of the mixed-VPA \eqref{iPraw} has a transverse momentum-dependent part augmented by a term proportional to $M_C^2\, g^{\mu\nu}$ (see Eq.~(52) in \cite{Chivukula:2011ng} for details). The latter arises due to the presence of massive states in the loop. Since the momentum-dependent part of the mixed-VPA \textit{is} transverse, the corresponding counterterm is also transverse in structure, as required by the original gauge-invariant Lagrangian.} This issue can be elegantly avoided by using the pinch technique as follows.

Taking the two non-Abelian vertices of the first diagram, defined in \eqref{VPAgam}, we may apply the decomposition \eqref{dec} to express their product as
\begin{equation}\label{G1G2}
\Gamma_1\Gamma_2 = \Gamma_1^F\Gamma_2^F+\Gamma_1^P\Gamma_2+\Gamma_2^P\Gamma_1-\Gamma_1^P\Gamma_2^P \ .
\end{equation}
The reason for expressing the product in this particular form is that the terms containing one factor of $\Gamma^P$ trigger Ward identities {\em a la} \eqref{WI}, albeit in a more complicated form, and this yields convenient simplifications. The pinch technique, therefore, allows the product of vertices to be rewritten as
\begin{equation}\label{G1G2fin}
\begin{split}
\Gamma_1\Gamma_2 = &\, \Gamma_1^F\Gamma_2^F + 2k_\mu(k+q)_\nu -2 \, (2k+q)_\mu(2k+q)_\nu + \left[k^2 - M_C^2\right]  g_{\mu\nu} + \left[(k+q)^2 - M_C^2\right]  g_{\mu\nu} \\
& -2\left[q^2 g_{\mu\nu}-q_\mu q_\nu\right] -2\left[(q^2-M_C^2) g_{\mu\nu}-q_\mu q_\nu\right] \ .
\end{split}
\end{equation}

Inserting \eqref{G1G2fin} back into \eqref{iPraw}, we can decompose the mixed-VPA
\begin{equation}\label{mVPAfin}
i\Pi^{ab}_{\mu\nu} (q) = i\hat{\Pi}^{ab}_{\mu\nu} (q) + i\Pi^{P\,ab}_{\mu\nu} (q) \ ,
\end{equation}
where
\begin{align}
i\hat{\Pi}^{ab}_{\mu\nu} (q) \equiv & \, g_s^2 \cot(2\theta_c)\, C_2(G)\, \delta^{ab} \mu^{4-d} \int \frac{d^d k}{(2\pi)^d} \left\{ \frac{\Gamma_1^F\Gamma_2^F-\frac{3}{2}(2k+q)_\mu(2k+q)_\nu}{\left[k^2-M_C^2+i\eta \right]\left[(k+q)^2-M_C^2+i\eta \right]} + \frac{(3-2d) g_{\mu\nu}}{k^2-M_C^2+i\eta} \right\} \ , \label{iPhat} \\
i\Pi^{P\,ab}_{\mu\nu} (q) \equiv & -2g_s^2 \cot(2\theta_c)\, C_2(G)\, \delta^{ab} \mu^{4-d} \int \frac{d^d k}{(2\pi)^d} \bigg\{ \frac{q^2 g_{\mu\nu}-q_\mu q_\nu}{\left[k^2-M_C^2+i\eta \right]\left[(k+q)^2-M_C^2+i\eta \right]} \notag \\
& \qquad \qquad \qquad \qquad \qquad \qquad \qquad \qquad \quad + \frac{(q^2-M_C^2) g_{\mu\nu}-q_\mu q_\nu}{\left[k^2-M_C^2+i\eta \right]\left[(k+q)^2-M_C^2+i\eta \right]} \bigg\} \ , \label{iPP}
\end{align}
and
\begin{equation}\label{GF12}
\Gamma_1^F\Gamma_2^F = d \, (2k+q)_\mu(2k+q)_\nu + 8 \brac{ q^2 g_{\mu\nu}-q_\mu q_\nu }\ .
\end{equation}

Explicit calculation of $i\hat{\Pi}^{ab}_{\mu\nu} (q)$ \eqref{iPhat} using standard methods in $d=4-2\epsilon$ reveals its transverse structure
\begin{equation}\label{iPhatfin}
i\hat{\Pi}^{ab}_{\mu\nu} (q) = \frac{i \alpha_s}{4\pi} \cot(2\theta_c) \, C_2(G) \delta^{ab} \left[\frac{43}{6}E+G(R_C)\right] \brac{ q^2g_{\mu\nu}-q_\mu q_\nu} \ ,
\end{equation}
with $\alpha_s = \dfrac{g_s^2}{4\pi}$, and where we have defined
\begin{align}
E \equiv &\, \frac{1}{\epsilon} - \gamma + \log 4\pi -\log \frac{M_C^2}{\mu^2} \label{E} \ , \\
G(R_C) \equiv &\, \frac{10}{9}\brac{6R_C+13}-\frac{1}{3}\brac{20R_C+43}\sqrt{4R_C-1} \, \text{arccot} \sqrt {4R_C-1} \ , \qquad R_C \equiv \frac{M_C^2}{q^2} \label{G} \ .
\end{align}
In \eqref{E}, $\gamma$ is the Euler-Mascheroni constant. Note that the function $G(R_C)$ in \eqref{G} is a finite well-defined function of $R_C$ that has the following values in the useful limiting cases
\begin{equation}\label{Gval}
G(\infty) = \frac{2}{3} \quad \text{(gluon on-shell)} \ , \qquad G(1) = \frac{190}{9} - \frac{7\sqrt3}{2}\pi \quad \text{(coloron on-shell)} \ .
\end{equation}
The expression \eqref{iPhatfin} should be kept as the gauge-invariant mixed-VPA contribution in this part of the calculation. Its transversality, as exposed by the pinch technique, exhibits the gauge-invariance of the theory, which is ordinarily obscured by its spontaneously broken non-Abelian nature. We will clarify below how to deal further with the leftover piece \eqref{iPP} properly.

Inserting \eqref{mVPAfin} into \eqref{iS}, we obtain
\begin{equation}\label{iSfin}
iS = i\hat{S} + iS^P \ .
\end{equation}
The $i\hat S$ structure, containing $i\hat{\Pi}^{ab}_{\mu\nu} (q)$ \eqref{iPhatfin}, forms the gauge-invariant mixed-VPA contribution to the gluon-initiated production of a coloron. Note that in the case of phenomenological interest to us, the external gluon legs will be on-shell; however, the fact that \eqref{iPhatfin} is transverse means that the second and third diagrams of Fig.~\ref{gaugeS} will vanish in this situation.\footnote{In case the gluons are off-shell their (UV-divergent) mixed-VPA is removed by the same counterterm as the non-mixed gluon VPA, as expected from gauge invariance. The latter must be also pinch-technique massaged into a transverse form, due to the presence of massive states in its loop. The second counterterm in the theory is used for the non-mixed coloron VPA renormalization, as explained in \cite{Chivukula:2011ng}.} Only the first diagram of Fig.~\ref{gaugeS}, where the mixed-VPA is within the external coloron line, will contribute to the amplitude
\begin{equation}\label{iShat}
i\hat S = \frac{g_s \alpha_s}{4\pi} \cot(2\theta_c) \, C_2(G) f^{amn} \left[\frac{43}{6}E+G(R_C)\right] \Big\{\Gamma^{\nu\alpha\beta}(-r,p,\bar{p}) + \dots \Big\} \ .
\end{equation}
where the $1/r^2$ term of the internal coloron propagator is cancelled by the $r^2 g^{\nu}_{\rho}$ part of \eqref{iPhatfin}, and the ellipses denote, collectively, the terms that can be discarded (including the $r^\nu r_\rho$ terms).\footnote{In the on-shell case, these terms are zero. In the case where the external gauge boson is off-shell, these terms may still be discarded, as they give rise to (pinched) structures which, in light of the generality of the pinch technique, need to be reassigned to the external vertices connected to that off-shell propagator and, hence, removed from the current discussion. This is sometimes referred to as the \textit{intrinsic} pinch technique \cite{Binosi:2009qm}.} 

\begin{figure}
\begin{center}
\includegraphics[width=.85\textwidth]{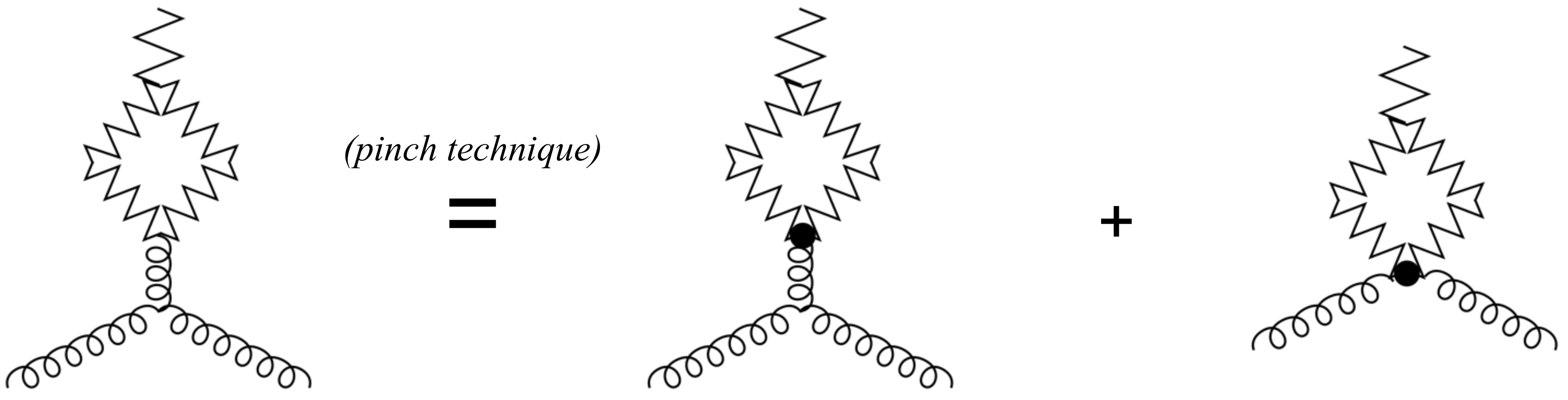}
\caption{Schematics of the decomposed non-Abelian mixed-VPA contribution to the coloron production amplitude, according to the pinch technique. The unpinched part contributes together with the remaining diagrams (Fig.~\ref{mVPA}) to $i\hat S$ \eqref{iShat}, whereas the pinched diagram, corresponding to $iS^P$ \eqref{iSP}, is added to the vertex structures sharing the same topology. Similar structures arise for the other two external leg permutations.}
\label{PTS}
\end{center}
\end{figure}

Of course, we still need to consider $iS^P$, containing the leftover piece $i\Pi^{P\,ab}_{\mu\nu} (q)$ of the mixed-VPA \eqref{iPP}. This may be written as
\begin{equation}\label{iSP}
\begin{split}
iS^P =&\, 2i \, g_s^3\cot(2\theta_c) \, C_2(G)f^{amn} \, \Gamma^{\nu \alpha\beta} (-r,p,\bar{p}) \, \mu^{4-d} \int \frac{d^d k}{(2\pi)^d} \bigg\{ \frac{1}{\left[k^2-M_C^2+i\eta \right]\left[(k+r)^2-M_C^2+i\eta \right]} \\
& \; + \frac{1}{\left[k^2-M_C^2+i\eta \right]\left[(k+p)^2-M_C^2+i\eta \right]} + \frac{1}{\left[k^2-M_C^2+i\eta \right]\left[(k+\bar p)^2-M_C^2+i\eta \right]} \bigg\} + \dots \ .
\end{split}
\end{equation}
Note the disappearance of the internal propagators (originally present in \eqref{iS}) in this expression, as they are cancelled against the numerators of \eqref{iPP}. It is of central importance to recognize that this cancellation of the internal propagators occurs {\em prior} to performing any loop-momentum integral. Therefore, the disappearance of the internal propagators generates pinched Feynman diagrams, as depicted in Fig.~\ref{PTS}, which possess an altered topology as compared with the original diagrams. We will see in the next section that these pinched structures need to be added to the vertex diagrams of similar form, and hence, should be removed altogether from the current discussion, leaving us only with the gauge-invariant mixed-VPA contribution \eqref{iShat}.

\subsection{Vertex Contributions {\em a la} Pinch Technique: Gauge Sector}

\begin{figure}
\begin{center}
\includegraphics[width=.65\textwidth]{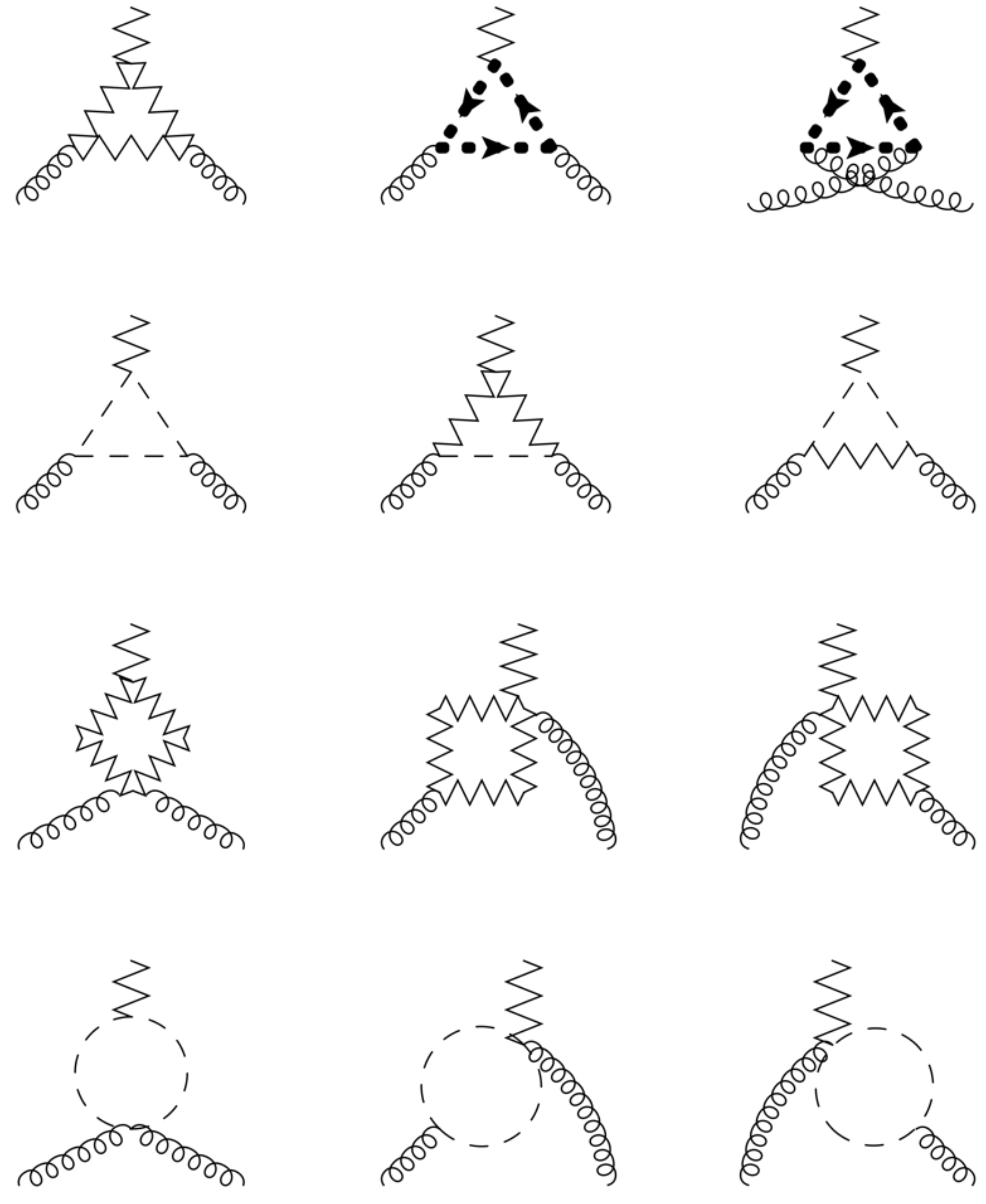}
\caption{The gauge sector one-loop vertex contributions to coloron production via gluon fusion. The particle definitions are as in Fig.~\ref{mVPA}. The last two diagrams on the second row are proportional to $M_C^2$ and are finite by powercounting. Those in the final row vanish, as they contain the product of symmetric and antisymmetric color factors.}
\label{gaugeV}
\end{center}
\end{figure}

The gauge sector one-loop vertex contributions are given by the diagrams of Fig.~\ref{gaugeV}, with the first diagram containing three non-Abelian vertices. By analogy with the previous section, let us define these as
\begin{equation}\label{Vgam}
\Gamma_1 \equiv \Gamma^{\nu\lambda \rho}(-r,k+\bar p,-(k-p)) \ , \quad \Gamma_2 \equiv \Gamma^{\alpha\mu}_{\quad \rho}(p,-k,k-p) \ , \quad \Gamma_3 \equiv \Gamma^{\beta}_{\; \lambda \mu}(\bar p,-(k+ \bar p),k) \ ,
\end{equation}
where $k$ is again the loop momentum, and we have used the definition \eqref{Gamma}. Employing the color product identity
\begin{equation}\label{fid}
f^{aed}f^{mdh}f^{nhe} = -i\,\text{Tr} F^a F^m F^n = \frac{1}{2} C_2(G) f^{amn} \ , \qquad (F^a)^{bc} \equiv -if^{abc} \ ,
\end{equation}
we get the following expression for the gauge vertex contributions
\begin{equation}\label{iV}
i V= i g_s^3 \cot (2\theta_c) \, C_2(G) f^{amn}\, \mu^{4-d} \int \frac{d^d k}{(2\pi)^d} \left \{ \frac{N^{\nu \alpha \beta}}{[k^2-M_C^2+i \eta] \, [(k-p)^2-M_C^2+i \eta] \, [(k+\bar{p})^2-M_C^2+i \eta]} +  B^{\nu \alpha \beta} \right \} \ , 
\end{equation}
with
\begin{align}
N^{\nu \alpha \beta} \equiv & \, \Gamma_1\Gamma_2\Gamma_3 -(k-p)^\nu \, k^\alpha \, (k+\bar{p})^\beta - (k+\bar{p})^\nu \, k^\beta \, (k-p)^\alpha + \frac{1}{2} (2k-p+\bar{p})^\nu \, (2k-p)^\alpha \, (2k+\bar{p})^\beta \notag \\
& \,- M_C^2 \, g^\alpha_{\rho} g^\beta_{\lambda} \, \Gamma_{1} -\frac{1}{2}M_C^2 \, g^{\alpha\beta} (2k-p+\bar p)^\nu \ , \label{N} \\
B^{\nu \alpha \beta} \equiv & \, \frac{9}{2} \bigg\{ \frac{r^\alpha \, g^{\beta \nu} - r^\beta \, g^{\alpha \nu}}{\left[k^2-M_C^2+i\eta \right]\left[(k+r)^2-M_C^2+i\eta \right]} + \frac{p^\nu \, g^{\alpha \beta} - p^\beta \, g^{\alpha \nu}}{\left[k^2-M_C^2+i\eta \right]\left[(k+p)^2-M_C^2+i\eta \right]} \notag \\
&\quad +\frac{\bar{p}^\alpha \, g^{\beta \nu} - \bar{p}^\nu \, g^{\alpha \beta}}{\left[k^2-M_C^2+i\eta \right]\left[(k+\bar p)^2-M_C^2+i\eta \right]} \bigg\} \label{B} \ .
\end{align}
Here, $N^{\nu \alpha \beta}$ is the summed Lorentz structure of the diagrams in the first two rows of Fig.~\ref{gaugeV}, while $B^{\nu \alpha \beta}$ is that of the third row. The diagrams in the last row of the same figure vanish due to the presence of the product of symmetric and antisymmetric color factors.

\begin{figure}
\begin{center}
\includegraphics[width=.85\textwidth]{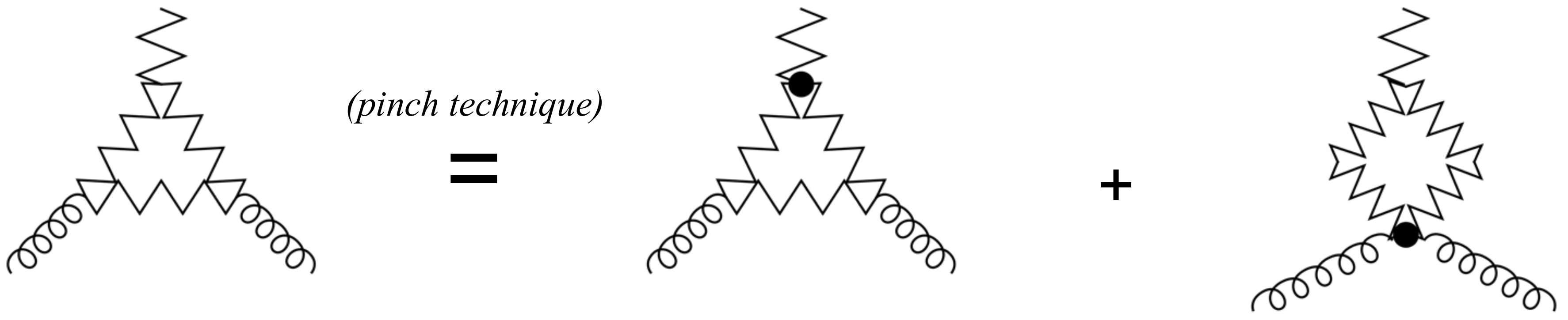}
\caption{Schematics of the decomposed non-Abelian vertex contribution to the coloron production amplitude, according to the pinch technique. The pinched structures, originating from $N^{\nu \alpha \beta}$ part of the amplitude \eqref{N}, are reassigned to $B^{\nu \alpha \beta}$ \eqref{B} due to their common topology, while the unpinched structures remain in $N^{\nu \alpha \beta}$. Similar structures arise for the other two vertex permutations.}
\label{PTV}
\end{center}
\end{figure}

At this stage, let us apply the pinch technique to the vertex contribution; namely, to the first diagram of Fig.~\ref{gaugeV}. Decomposing its non-Abelian vertices \eqref{Vgam} according to \eqref{dec} as before, we may write for their product
\begin{equation}\label{G1G2G3}
\Gamma_1\Gamma_2\Gamma_3 = \Gamma_1^F\Gamma_2^F\Gamma_3^F+\Gamma_1^P\Gamma_2\Gamma_3+\Gamma_1\Gamma_2^P\Gamma_3+\Gamma_1\Gamma_2\Gamma_3^P-\Gamma_1^P\Gamma_2^P\Gamma_3-\Gamma_1^P\Gamma_2\Gamma_3^P-\Gamma_1\Gamma_2^P\Gamma_3^P+\Gamma_1^P\Gamma_2^P\Gamma_3^P \ .
\end{equation}
Terms containing one or two factors of $\Gamma^P$ can be expressed in convenient forms using more complicated Ward identities {\em a la} \eqref{WI}, as in the previous section. This allows for significant simplifications, and we obtain for the final pinch-technique rearranged product
\begin{equation}\label{G1G2G3fin}
\Gamma_1\Gamma_2\Gamma_3 = \Gamma^N_{123} + \Gamma^B_{123} \ ,
\end{equation}
with
\begin{align}
\Gamma^N_{123} \equiv &\, \Gamma_1^F\Gamma_2^F\Gamma_3^F + (k-p)^\nu \, k^\alpha \, (k+\bar{p})^\beta + (k+\bar{p})^\nu \, k^\beta \, (k-p)^\alpha -2 \, (2k-p+\bar{p})^\nu \, (2k-p)^\alpha \, (2k+\bar{p})^\beta \notag \\
& + M_C^2 \left[ (3k-p+2\bar{p})^\nu g^{\alpha \beta} - (k+p+2\bar{p})^\alpha g^{ \beta \nu} - (k-2p-\bar{p})^\beta g^{\alpha \nu} \right] + \dots \ , \label{GN} \\
\Gamma^B_{123} \equiv&\, [k^2 - M_C^2] \left[ -2 g^{\alpha \beta} (p-\bar{p})^\nu + g^{\alpha \nu} (k-\bar{p})^\beta +g^{\beta \nu} (k+p)^\alpha \right] \notag \\
& + [(k+\bar{p})^2 - M_C^2] \left[ g^{\alpha \beta} (k+p+2\bar{p})^\nu + g^{\alpha \nu} (k+2\bar{p})^\beta - 2g^{\beta \nu} (p+2\bar{p})^\alpha \right] \notag \\
& + [(k-p)^2 - M_C^2] \left[ g^{\alpha \beta} (k-2p-\bar{p})^\nu + 2g^{\alpha \nu} (2p+\bar{p})^\beta + g^{\beta \nu} (k-2p)^\alpha \right]\label{GB} \ ,
\end{align}
where ellipses again denote the terms that may be discarded (see footnote~12). The terms in \eqref{GB} cancel an internal propagator, furnishing pinched diagrams with the same topology as those in the third row of Fig.~\ref{gaugeV}. Consequently, certain structures are reassigned from $N^{\nu \alpha \beta}$ \eqref{N} to $B^{\nu \alpha \beta}$ \eqref{B}, as prescribed by the pinch technique. This has been diagrammatically illustrated in Fig.~\ref{PTV}.

There is, however, one more contribution that must be considered in this discussion. Recalling the pinched structures from the mixed-VPA,  $iS_P$ \eqref{iSP}, we observe that they share the same topology as the diagrams containing $B^{\nu \alpha \beta}$ \eqref{B}, and should be added to them as well. This turns out to be an essential operation, in order to recover a consistent QED-like Ward identity. We obtain the full pinch-technique constructed vertex contribution
\begin{align}
i\hat{V} =&\, iV + iS^P \label{iVhat} \\
=&\, i g_s^3 \cot (2\theta_c) \, C_2(G) f^{amn}\, \mu^{4-d} \int \frac{d^d k}{(2\pi)^d} \left \{ \frac{\hat{N}^{\nu \alpha \beta}}{[k^2-M_C^2+i \eta] \, [(k-p)^2-M_C^2+i \eta] \, [(k+\bar{p})^2-M_C^2+i \eta]} + \hat{B}^{\nu \alpha \beta} \right \} \notag \ , 
\end{align}
where
\begin{align}
\hat{N}^{\nu \alpha \beta} \equiv & \, \Gamma^F_1\Gamma^F_2\Gamma^F_3 - \frac{3}{2} (2k-p+\bar{p})^\nu \, (2k-p)^\alpha \, (2k+\bar{p})^\beta + \dots \ ,\notag \\
\hat{B}^{\nu \alpha \beta} \equiv & \, 8 \bigg\{ \frac{r^\alpha \, g^{\beta \nu} - r^\beta \, g^{\alpha \nu}}{\left[k^2-M_C^2+i\eta \right]\left[(k+r)^2-M_C^2+i\eta \right]} + \frac{p^\nu \, g^{\alpha \beta} - p^\beta \, g^{\alpha \nu}}{\left[k^2-M_C^2+i\eta \right]\left[(k+p)^2-M_C^2+i\eta \right]} \notag \\
&\quad +\frac{\bar{p}^\alpha \, g^{\beta \nu} - \bar{p}^\nu \, g^{\alpha \beta}}{\left[k^2-M_C^2+i\eta \right]\left[(k+\bar p)^2-M_C^2+i\eta \right]} \bigg\} \label{Bp} \ ,
\end{align}
and
\begin{equation}\label{GF123}
\begin{split}
\Gamma_1^F\Gamma_2^F\Gamma_3^F =&\, d \, (2k-p+\bar{p})^\nu \, (2k-p)^\alpha \, (2k+\bar{p})^\beta + 8M_C^2\brac{k^\alpha g^{\beta\nu} + k^\beta g^{\alpha\nu}-k^\nu g^{\alpha\beta}} \\
&\,+8\left[ (2k-p+\bar p)^\nu p^\beta \bar{p}^\alpha - (2k-p)^\alpha r^\beta \bar{p}^\nu - (2k+\bar p)^\beta r^\alpha p^\nu+ r^\beta p^\nu \bar{p}^\alpha - r^\alpha p^\beta \bar{p}^\nu \right] + \dots \ .
\end{split}\end{equation}

The coefficient of the divergent part of \eqref{iVhat} can be calculated for the general case in a closed form. In $d=4-2\epsilon$ dimensions, we obtain
\begin{equation}\label{iVhatinf}
i\hat V_{\text{infinite}} = -\frac{g_s \alpha_s}{4\pi} \cot(2\theta_c) \, C_2(G) f^{amn} \brac{ \frac{43}{6}E } \Gamma^{\nu\alpha\beta}(-r,p,\bar{p}) \ ,
\end{equation}
with $E$ given in \eqref{E} and using the definition in \eqref{Gamma}. The remaining finite piece of \eqref{iVhat} can be evaluated for the on-shell external gauge bosons \eqref{EOM} using standard methods, and leads to
\begin{equation}\label{iVhatfin}
\begin{split}
i\hat V_{\text{finite}} =&\, -\frac{g_s \alpha_s}{4\pi} \cot(2\theta_c) \, C_2(G) f^{amn} \bigg[ \brac{\frac{380}{9}-7\sqrt3 \pi} \brac{\bar p^\alpha g^{\beta \nu} -p^\beta g^{\alpha \nu}} + \frac{5-3\sqrt3 \pi +5\pi^2}{18} \brac{p-\bar p}^\nu g^{\alpha \beta} \\
&\qquad +\frac{5\brac{75-12\sqrt3 \pi - \pi^2}}{9\,M_C^2} \brac{p-\bar p}^\nu p^\beta \bar p^\alpha\bigg] \ .
\end{split}
\end{equation}

At this point, let us pause and reflect on the implemented procedure in the gauge sector of the theory. Applying the pinch technique, we have shuffled around terms within the vertex contribution and classified them into two intrinsically irreducible topologies. In addition, we have removed certain structures from the mixed-VPA contributions and reassigned them to vertices with similar topology. The full generality of this procedure should be appreciated, as it has been carried out prior to performing any loop-momentum integration. Moreover, this process has enabled us to construct a transverse mixed-VPA, manifestly exhibiting the gauge invariance of the theory. Furthermore, one can show that, unlike the conventional one-loop vertex \eqref{iV}, the pinch-technique constructed $i\hat{V}$ \eqref{iVhat} satisfies a simple QED-like Ward identity, as illustrated in \cite{Binosi:2009qm}.

Interestingly, one observes that the UV divergence of the vertex \eqref{iVhatinf} exactly cancels that of the mixed-VPA \eqref{iShat}. This cancellation is, however, not a coincidence, but a consequence of the above mentioned recovered QED-like Ward identity. The amplitude for coloron production via gluon fusion is, thus, finite!\footnote{This is expected, as there are no available counterterms possessing the form of this one-loop amplitude in the Lagrangian.}

\subsection{Vertex Contributions: Matter Sector}

\begin{figure}
\begin{center}
\includegraphics[width=.5\textwidth]{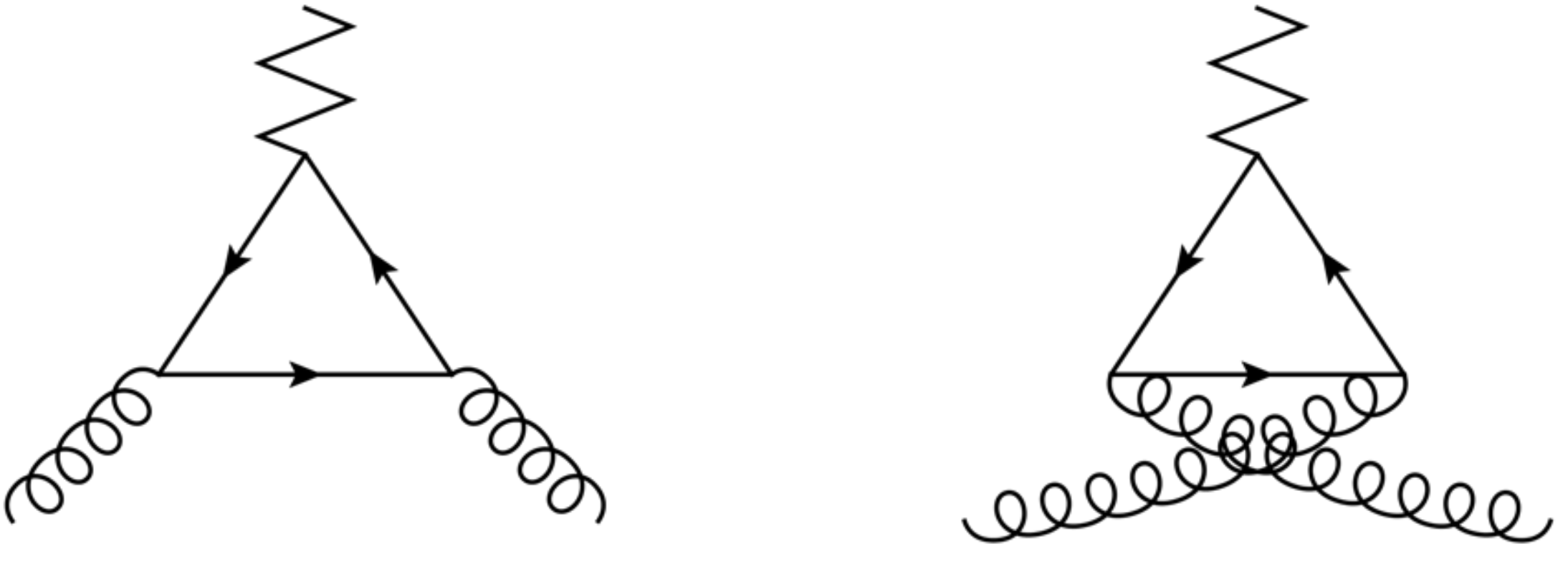}
\caption{Triangle diagrams with the chiral massless quarks in the loop.}
\label{qV}
\end{center}
\end{figure}

Having explored the structure of the gauge sector of the theory, we now examine the contributions to the production amplitude originating from fermion loops. The vertex contributions with quarks running in the loop are depicted in Fig.~\ref{qV}. The gluon coupling is vectorial, whereas the quark coupling to the coloron is assumed to be chiral in general (c.f. \eqref{Lferm}). It is well known that these triangle diagrams can generally be anomalous; thereby, endangering the consistency of the theory.

\begin{figure}
\begin{center}
\includegraphics[width=.5\textwidth]{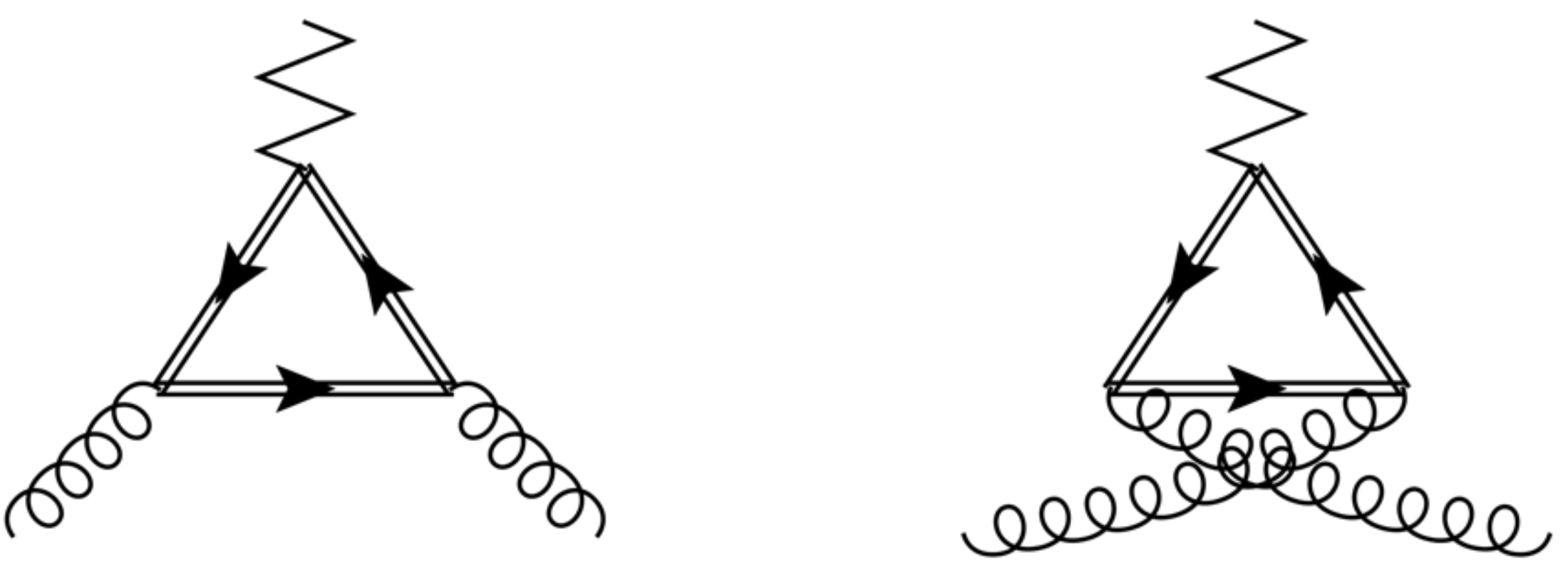}
\caption{Triangle diagrams with massive spectator fermions in the loop. The spectators are represented by the continuous double lines. As described in the text, each spectator has chirality opposite to that of its corresponding ordinary quark.}
\label{QV}
\end{center}
\end{figure}

In order to preserve formal consistency, the full coloron theory must, therefore, include additional fermions whose contributions cancel the potential anomalies of the ordinary quarks. We will estimate these contributions by using the simplest set of \textit{spectator} fermions that cancels the anomalies. This consists of a heavy replica of each quark flavor with the opposite chirality. For simplicity, we assume a degenerate mass scale for all the spectator flavors, $M_Q$, above the lower bound presently set by collider limits, while the ordinary quarks are set to be massless (see Appendix~\ref{FR} for the Feynman rules). The triangle diagrams containing the spectators are illustrated in Fig.~\ref{QV}.

The fermionic one-loop vertex contribution, now containing both the massless quarks and the heavy spectators, is of the form
\begin{equation}\label{iVfraw}
iV_{\text{ferm}} = iV_{\text{ferm}}^V + iV_{\text{ferm}}^A \ ,
\end{equation}
where\footnote{As in the case with the quarks, we assume that the coloron coupling to spectators is flavor-universal -- the generalization to the flavor-dependent case is straightforward (see footnote~7).}
\begin{align}
iV_{\text{ferm}}^V \equiv \, & -\frac{ig_s^3 (r_L+r_R)}{4}\, N_f f^{amn}\, \mu^{4-d} \int \frac{d^d k}{(2\pi)^d} \label{iVfV} \\
& \Bigg\{  \frac{\text{Tr} \left[ \gamma^\nu (\slashed{k}+\bar{\slashed{p}}) \gamma^\beta \slashed{k} \gamma^\alpha (\slashed{k}-\slashed{p}) \right]}{\left[ (k+\bar{p})^2 + i\eta \right] \left[ k^2 + i\eta \right] \left[ (k-p)^2 + i\eta \right]} 
+\frac{\text{Tr} \left[ \gamma^\nu (\slashed{k}+\bar{\slashed{p}}+M_Q) \gamma^\beta (\slashed{k}+M_Q) \gamma^\alpha (\slashed{k}-\slashed{p}+M_Q) \right]}{[ (k+\bar{p})^2 -M_Q^2 + i\eta] \,[ k^2 -M_Q^2 + i\eta] \,[ (k-p)^2 -M_Q^2 + i\eta ]} \Bigg \} \notag \ , \\
iV_{\text{ferm}}^A \equiv \, & -\frac{g_s^3 (r_L-r_R)}{4}\, N_f d^{amn}\, \mu^{4-d} \int \frac{d^d k}{(2\pi)^d}  \label{iVfA} \\
& \Bigg\{  \frac{\text{Tr} \left[ \gamma^\nu\gamma_5 (\slashed{k}+\bar{\slashed{p}}) \gamma^\beta \slashed{k} \gamma^\alpha (\slashed{k}-\slashed{p}) \right]}{\left[ (k+\bar{p})^2 + i\eta \right] \left[ k^2 + i\eta \right] \left[ (k-p)^2 + i\eta \right]} 
-\frac{\text{Tr} \left[ \gamma^\nu\gamma_5 (\slashed{k}+\bar{\slashed{p}}+M_Q) \gamma^\beta (\slashed{k}+M_Q) \gamma^\alpha (\slashed{k}-\slashed{p}+M_Q) \right]}{[ (k+\bar{p})^2 -M_Q^2 + i\eta] \,[ k^2 -M_Q^2 + i\eta] \,[ (k-p)^2 -M_Q^2 + i\eta ]} \Bigg \} \notag \ .
\end{align}
Here, $iV_{\text{ferm}}^V$ denotes the vectorial part of this amplitude, whereas $iV_{\text{ferm}}^A$ is the axial contribution. The $N_f$ is the number of fermion flavors in the loop, which is chosen to be equal for the quarks and the spectators.  In addition, we have employed the definitions \eqref{PLPR} and \eqref{rLrR}, along with the fundamental color trace
\begin{equation}\label{tid}
\text{Tr}\left[ t^a t^m t^n \right] = \frac{1}{4} \brac{d^{amn} + if^{amn}} \ ,
\end{equation}
with $d^{amn}$ symmetric under the interchange of any two indices.

By dotting the external gauge bosons' momenta, it can be demonstrated that the one-loop fermionic vertex \eqref{iVfraw} satisfies the usual Ward identities for massive and massless fermions, with the cancellation of the anomaly occurring specifically between the quarks and the spectators.\footnote{Defining $\gamma_5$ in $d$ dimensions is ambiguous. In order to compute the anomaly and demonstrate its cancellation in dimensional regularization, one {\em must} employ the Veltman-'t Hooft definition of $\gamma_5$, as the `naive' prescription will not reproduce the anomaly. Alternatively, one might choose a different regularization scheme, such as the Pauli-Villars, to calculate the anomaly and exhibit its cancellation.} This proves that the theory is anomaly-free once the spectator fermions are included.

Given the anomaly-free nature of \eqref{iVfraw}, one may compute this one-loop fermionic vertex unambiguously. As was the case with the gauge sector, the divergent contribution can be calculated for the general case in a closed form. Again using standard methods in $d=4-2\epsilon$ dimensions, one obtains
\begin{equation}\label{iVfinf}
iV_{\text{ferm, infinite}} = \frac{g_s \alpha_s}{4\pi}(r_L+r_R)\, N_f f^{amn} \brac{ \frac{2}{3}E } \Gamma^{\nu\alpha\beta}(-r,p,\bar{p}) \ ,
\end{equation}
where $E$ is given in \eqref{E} and definition \eqref{Gamma} is implied. 

Defining the spectator to coloron mass ratio as
\begin{equation}\label{R}
R \equiv \frac{M_Q^2}{M_C^2} \ ,
\end{equation}
the finite piece of \eqref{iVfraw} can be evaluated by setting the external gauge bosons on-shell \eqref{EOM}; it yields the expression\footnote{Technically, the fermionic vertex contains also a finite anomaly-free axial contribution, a remnant of the non-exact cancellation among the quarks and spectators in \eqref{iVfA}, due to their non-degenerate masses. By Yang's theorem \cite{Yang:1950rg}, this axial part, being proportional to the symmetric color factor $d^{amn}$, does not contribute to the cross section for the on-shell external states, and is omitted henceforth.}
\begin{equation}\label{iVffin}
iV_{\text{ferm, finite}} = \frac{g_s \alpha_s}{4\pi} (r_L+r_R)\, N_f f^{amn}\left[ F(R) \brac{\bar p^\alpha g^{\beta \nu} -p^\beta g^{\alpha \nu}} + F^\prime(R)\brac{p-\bar p}^\nu g^{\alpha \beta} +\frac{F^{\prime\prime}(R)}{M_C^2} \brac{p-\bar p}^\nu p^\beta \bar p^\alpha\right] \ ,
\end{equation}
with the functions
\begin{equation}\label{Fs}
\begin{split}
F(R) \equiv &\, \frac{2}{9} \left[ 10-3\pi i +12R-3\log R-6(2R+1)\sqrt{4R-1}\,\text{arccot}\sqrt{4R-1} \right] \ , \\
F^\prime(R) \equiv &\, \frac{1}{9} \left[ 7-3\pi i -24R-3\log R+6(4R-1)^{3/2}\,\text{arccot}\sqrt{4R-1} \right] +F_A(R) \ , \\
F^{\prime\prime}(R) \equiv &\, \frac{2}{3} \left[ 1+12R-12R\sqrt{4R-1}\,\text{arccot}\sqrt{4R-1} \right] -2F_A(R) \ , \\
F_A(R) \equiv&\, R\left[\text{Li}_2\brac{\frac{2}{1+\sqrt{1-4R}}} +\text{Li}_2\brac{\frac{2}{1-\sqrt{1-4R}}}\right] \ .
\end{split}
\end{equation}
Here, $\text{Li}_2(z)$ is the Jonqui\`ere's dilogarithm. The functions in \eqref{Fs} are constructed for the $R>\frac{1}{4}$ region, which defines the kinematic threshold for coloron decay into a pair of the spectator fermions.\footnote{In the current analysis, we restrict our attention to the $R>\frac{1}{4}$ regime, as a study of the direct production of the spectator fermions, although very interesting, is outside the scope of this treatment.}

\subsection{Vacuum Polarization Amplitudes: Matter Sector}

\begin{figure}
\begin{center}
\includegraphics[width=.8\textwidth]{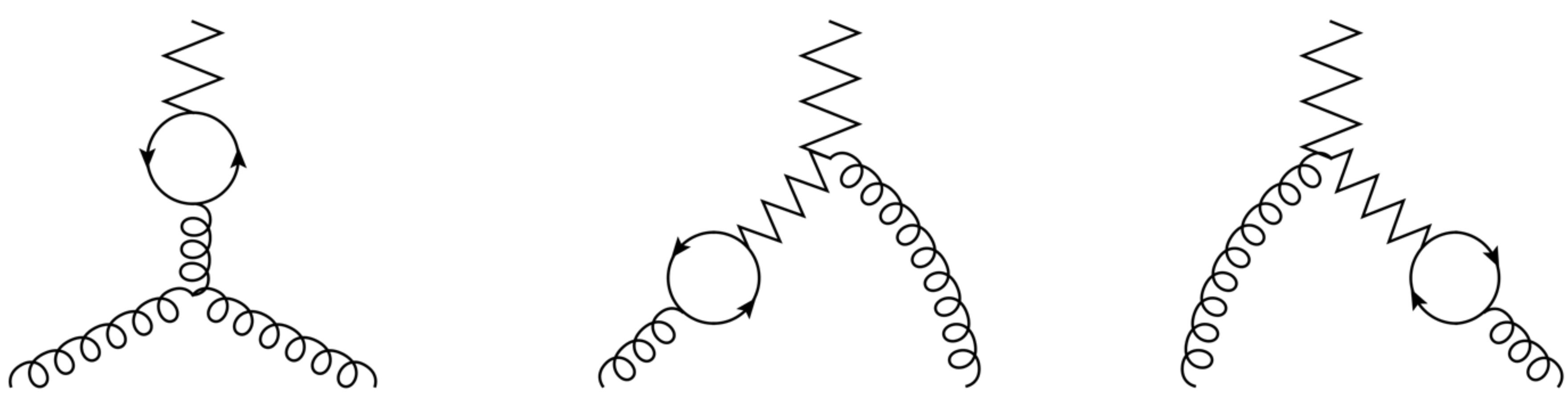}
\includegraphics[width=.8\textwidth]{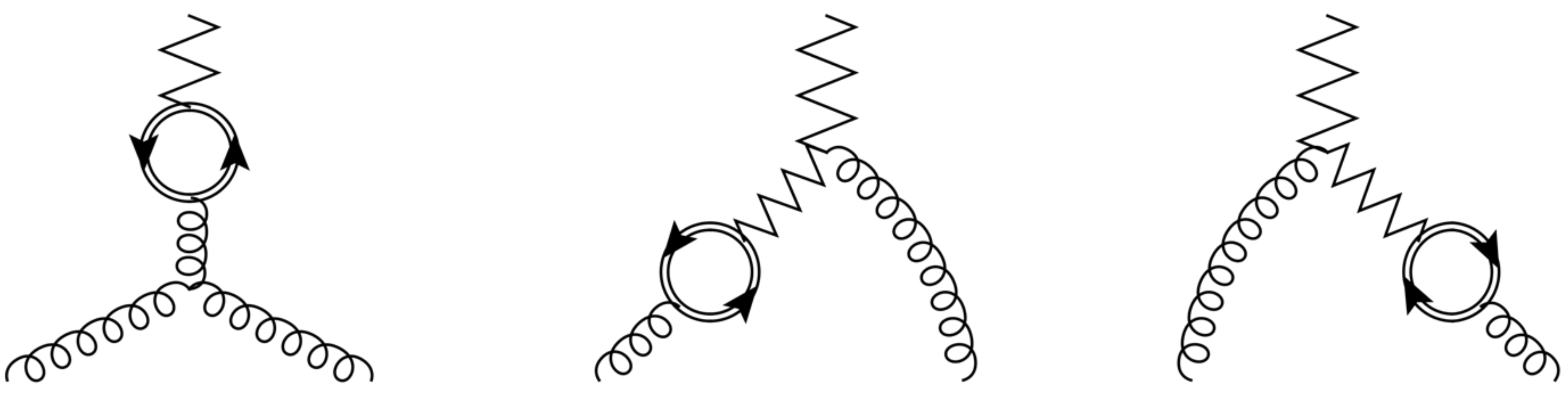}
\caption{Gluon fusion to coloron via the mixed-VPA containing fermions in the loop. The top row illustrates the quark contribution, while the bottom row represents that of the spectators.}
\label{fermS}
\end{center}
\end{figure}

The final piece of this calculation involves computing the fermionic mixed-VPA contributions to the production amplitude. The quark and spectator one-loop diagrams are illustrated in Fig.~\ref{fermS}. In analogy with the gauge sector \eqref{iS}, these contributions may be written as
\begin{equation}\label{iSf}
iS_{\text{ferm}} = g_s \left\{ f^{dmn} \Gamma^{\rho\alpha\beta}(-r,p,\bar{p}) \, \frac{\Pi^{ad}_{\text{ferm }\nu\rho} (r)}{r^2} + f^{adn} \Gamma^{\nu\rho\beta}(-r,p,\bar{p}) \, \frac{\Pi^{md}_{\text{ferm }\alpha\rho} (p)}{p^2-M_C^2} + f^{amd} \Gamma^{\nu\alpha\rho}(-r,p,\bar{p}) \,\frac{\Pi^{nd}_{\text{ferm }\beta\rho} (\bar{p})}{\bar{p}^2-M_C^2} \right\} \ ,
\end{equation}
with
\begin{equation}\label{iPf}
i\Pi^{ab}_{\text{ferm }\mu\nu} (q) = -\frac{g_s^2 \brac{r_L+r_R}}{4} \, N_f\delta^{ab} \mu^{4-d} \int \frac{d^d k}{(2\pi)^d} \left\{ \frac{\text{Tr}\left[ \gamma_\mu \slashed{k} \gamma_\nu (\slashed{k}+\slashed{q}) \right]}{\left[k^2+i\eta \right]\left[(k+q)^2+i\eta \right]} + \frac{\text{Tr}\left[ \gamma_\mu (\slashed{k}+M_Q) \gamma_\nu (\slashed{k}+\slashed{q}+M_Q) \right]}{[k^2-M_Q^2+i\eta]\,[(k+q)^2-M_Q^2+i\eta]} \right\} \ .
\end{equation}
The expression $i\Pi^{ab}_{\text{ferm }\mu\nu} (q)$ is the fermionic mixed-VPA, containing both the quarks and the spectators, which can be evaluated using standard methods. It yields the expected transverse structure
\begin{equation}\label{iPffin}
i\Pi^{ab}_{\text{ferm }\mu\nu} (q) = -\frac{i \alpha_s \brac{r_L+r_R}}{4\pi} \, N_f\delta^{ab} \left[\frac{2}{3}E+\frac{1}{2}F(R_Q)\right] \brac{ q^2g_{\mu\nu}-q_\mu q_\nu} \ ,
\end{equation}
where $F(R_Q)$ is given in \eqref{Fs} with the definition (c.f. $R_C$ in \eqref{G})
\begin{equation} \label{RQ}
R_Q \equiv \frac{M_Q^2}{q^2} \ .
\end{equation}

Inserting the fermionic mixed-VPA \eqref{iPffin} into \eqref{iSf}, and recalling that the on-shell external gluon legs do not contribute (see the explanation below \eqref{iSfin}), we obtain the following amplitude for the portion of the gluon-gluon to coloron process that is facilitated by the gauge-invariant, fermionic mixed-VPA
\begin{equation}\label{iSffin}
iS_{\text{ferm}} = -\frac{g_s \alpha_s}{4\pi} \brac{r_L+r_R} N_f f^{amn} \left[\frac{2}{3}E+\frac{1}{2}F(R_Q)\right] \Big\{\Gamma^{\nu\alpha\beta}(-r,p,\bar{p}) + \dots \Big\} \ ,
\end{equation}
where ellipses denote terms that may be discarded, as before (see below \eqref{iShat}).

It is evident, once more, that the UV divergences of the fermionic one-loop vertices \eqref{iVfinf} and the mixed-VPA contributions \eqref{iSffin} cancel one-another, yielding a finite amplitude. Furthermore, a comparison between \eqref{iShat}, \eqref{iVhatfin}, \eqref{iVffin}, and \eqref{iSffin} establishes that the matter sector and the pinch-technique treated gauge sector consistently share the same formal structure, and may be readily combined. By virtue of the pinch technique, we have been able to make the gauge-invariant nature of the gauge sector transparent, which mimics the matter sector; a fact that, otherwise, might not have been easy to reveal.

\subsection{Production Amplitude of Coloron Via Gluon-Fusion}

We now possess all the pieces of the puzzle for constructing the leading order amplitude of the gluon-initiated coloron production \eqref{iM}, which now takes the form
\begin{equation}\label{iMpt}
i\mathcal{M}_{gg\to C} = (i\hat S + i\hat V + iS_{\text{ferm}} + iV_{\text{ferm}}) \times \varepsilon^{* a \, (\lambda_{C})}_{C \, \nu} (r) \, \varepsilon^{m \, (\lambda_{g_1})}_{g \, \alpha} (p) \, \varepsilon^{n \, (\lambda_{g_2})}_{g \, \beta} (\bar{p}) \ .
\end{equation}
Applying the on-shell identities \eqref{EOM}, we note that $R_C=1$ (see \eqref{G}) and $R_Q=R$ (see \eqref{RQ} and \eqref{R}). We may then insert the corresponding on-shell expressions of \eqref{iShat}, \eqref{iVhatinf}, \eqref{iVhatfin}, \eqref{iVfinf}, \eqref{iVffin}, and \eqref{iSffin} into \eqref{iMpt} to obtain the final form of the amplitude, containing both matter and gauge contributions
\begin{equation} \label{iMfin}
i\mathcal{M}_{gg\to C} = \frac{\alpha_s}{4\pi} \brac{T_G + T_M} i\mathcal{T} \ ,
\end{equation}
with
\begin{align}
T_G \equiv& \, \cot (2\theta_c) \, C_2(G)\, \frac{5}{18} \brac{75-12\sqrt3 \pi -\pi^2} \label{TV} \ , \\
T_M \equiv& \, -\frac{1}{2} \brac{r_L+r_R} N_f \, F^{\prime\prime}(R) \label{TVp} \ , \\
i\mathcal{T} \equiv& \, g_s\, f^{amn}  \left[g^{\alpha \beta}-\frac{2p^\beta \bar{p}^\alpha}{M_C^2}\right] (p-\bar{p})^\nu \, \times \varepsilon^{* a \, (\lambda_{C})}_{C \, \nu} (r) \, \varepsilon^{m \, (\lambda_{g_1})}_{g \, \alpha} (p) \, \varepsilon^{n \, (\lambda_{g_2})}_{g \, \beta} (\bar{p}) \label{iMV} \ ,
\end{align}
and $F^{\prime\prime}(R)$ is given in \eqref{Fs}. The quantities $T_G$ and $T_M$ correspond to the gauge and matter sectors, respectively. The production amplitude \eqref{iMfin} is manifestly finite, as anticipated, with all the UV divergences canceling among the vertex and mixed-VPA contributions in their corresponding sectors, as enforced by gauge invariance.

\section{Gluon-Fusion Coloron Production Rates at Hadron Colliders}
\label{GGCHadron}

\subsection{Production Cross Section of Coloron Via Gluon-Fusion}

The leading order partonic cross section for coloron production via gluon fusion is given by
\begin{equation}\label{CSpart}
\hat{\sigma}_{gg\to C} = \frac{\pi}{\hat{s}}\, \overline{| \mathcal{M}_{gg\to C}|^2} \,\delta(\hat s-M_C^2) = \frac{\pi}{\hat{s}^2}\, \overline{| \mathcal{M}_{gg\to C}|^2} \,\delta(1-\chi) \ ,
\end{equation}
with $\hat s$ the partonic CM energy, and $\overline{| \mathcal{M}_{gg\to C}|^2}$ the squared amplitude \eqref{iMfin}, averaged over initial and summed over final colors and spins. In addition, we have defined the parametrization
\begin{equation}\label{chi}
\chi \equiv \frac{M_C^2}{\hat s} \ .
\end{equation}

In $d=4-2\epsilon$ dimensions, $\overline{| \mathcal{M}_{gg\to C}|^2}$ takes the form
\begin{equation}\label{M2}
\overline{| \mathcal{M}_{gg\to C}|^2} = \brac{\frac{1}{\text{dim} (G)}}^2 \brac{\frac{1}{2(1-\epsilon)}}^2 \sum_{\text{spin \& color}} | \mathcal{M}_{gg\to C}|^2 = \frac{\alpha_s^3}{4\pi} \, \frac{C_2(G)}{2 \, \text{dim}(G)} \, |T_G + T_M|^2 \, \hat s \ ,
\end{equation}
where dim$(G)=8$ is the dimension of the adjoint representation, and the gluon polarization is modified in $d$~dimensions. In evaluating \eqref{M2}, we used the color identity $(f^{amn})^2 = \text{dim}(G) \cdot C_2(G)$, which follows from \eqref{ffid}. 

Inserting \eqref{M2} into \eqref{CSpart}, the partonic cross section for the process takes the form
\begin{equation}\label{CSpartfin}
\hat{\sigma}_{gg\to C} = \frac{\alpha_s^3}{\hat{s}} \, D \, |T|^2\, \delta(1-\chi) \ ,
\end{equation}
where we have defined
\begin{align}
D \equiv&\, \frac{C_2(G)}{8\, \text{dim}(G)} \label{D} \ , \\
T \equiv&\, T_G + T_M = \cot (2\theta_c) \, C_2(G)\, \frac{5}{18} \brac{75-12\sqrt3 \pi -\pi^2} + N_f \, F^{\prime\prime}(R) \, H(\theta_c) \label{T}\ .
\end{align}
The function $F^{\prime\prime}(R)$ is given in \eqref{Fs}, while $H(\theta_c)$ is determined by the chiral couplings of the quarks to the coloron, which depend on how we choose to make the quarks transform under the original $SU(3)_{1c} \times SU(3)_{2c}$ symmetry,
\begin{equation}\label{H} \begin{split}
H(\theta_c) =
\begin{cases}
\tan\theta_c &\quad r_L=r_R=-\tan\theta_c \\
-\cot(2\theta_c)&\quad r_L\neq r_R \\
-\cot\theta_c &\quad r_L=r_R=\cot\theta_c
\end{cases} \ .
\end{split} \end{equation}

The full leading order hadronic cross section for coloron production via gluon fusion at the LHC is determined by convolving the partonic cross section \eqref{CSpartfin} with the gluon parton distribution functions (PDF) within the protons
\begin{equation}\label{CS}
\sigma_{gg\to C} = \int dx_1 \int dx_2 \, f_g(x_1,\mu_F)f_g(x_2,\mu_F) \, \hat{\sigma}_{gg\to C} \ ,
\end{equation}
where $f_g(x_i,\mu_F)$ is the PDF of the $i^{th}$ gluon, $x_i$ its carried momentum fraction, and $\mu_F$ is the factorization scale. Taking the collision axis to be the 3-axis, the four-momenta
of the gluons are
\begin{equation} \label{4mom}
p=\frac{\sqrt{s}}{2}\left(x_1,0,0,x_1\right)\ ,\qquad
\bar{p}=\frac{\sqrt{s}}{2}\left(x_2,0,0,-x_2\right) \ ,
\end{equation}
with $s$ the CM energy of the colliding protons. This leads to (c.f. \eqref{chi})
\begin{equation}\label{x1x2}
\hat{s}=x_1x_2\, s \ , \qquad
\chi = \frac{M_C^2}{ x_1 x_2 \, s} \ .
\end{equation}
Trading the partonic CM energy, $\hat s$, for the hadronic one by means of \eqref{x1x2}, we obtain the final production cross section
\begin{equation}\label{CSfin}
\sigma_{gg\to C} = \frac{\alpha_s^3}{s} \, D \int \frac{dx_1}{x_1} \int \frac{dx_2}{x_2} \, f_g(x_1,\mu_F)f_g(x_2,\mu_F) \, |T|^2\, \delta(1-\chi) \ ,
\end{equation}
with $D$ and $T$ as defined in \eqref{D} and \eqref{T}, respectively.

Expression \eqref{CSfin} establishes the leading order cross section for the coloron production via gluon-fusion at the LHC. It has a formal dependence on the ratio of the spectator and coloron masses, as parametrized by $R$ \eqref{R}. As the leading order expression for the gluon fusion amplitude, it is also expected to be highly dependent on the factorization scale, $\mu_F$. As we will see, however, this overall contribution is very small when compared to the quark annhilation contributions, so this dependence is of little phenomenological consequence. This concludes our formal treatment of the gluon-initiated coloron production.

\subsection{Hadron Collider Production Rates from Gluon-Fusion}
\label{subsec:hadron}

With the formal expression for the production cross section at our disposal, we now present some numerical results. For this purpose, we employ CT10 \cite{Lai:2010vv} to evaluate the relevant PDFs. Furthermore, we note that the coloron coupling contains, in addition to the strong coupling constant, the mixing angle ($\theta_c$), which is a free parameter (see the Feynman rules in Appendix~\ref{FR}). We will define $\theta_c$ by identifying $g_s$ in (\ref{gLgR}) with the value determined by $\alpha_s(M_Z)$.
In the cross section, therefore, we make a distinction between the values of the renormalized strong coupling ($\alpha_s(\mu)$) associated with the gluons and those originating from the colorons. Setting the renormalization scale at the coloron mass, we evaluate the renormalized gluon strong coupling at $M_C$, while that associated with the coloron is chosen to be evaluated at the $Z$-pole mass, $M_Z$. Also, the PDF factorization scale has been set equal to the coloron mass, $\mu_F = M_C$. 

\begin{figure}
\begin{center}
\includegraphics[width=.329\textwidth]{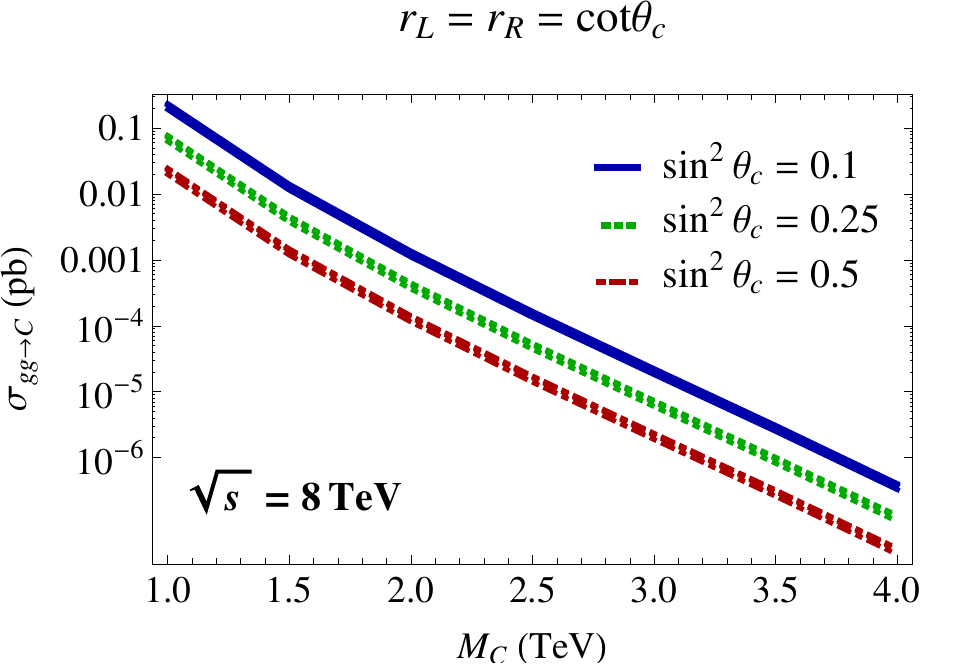}
\includegraphics[width=.329\textwidth]{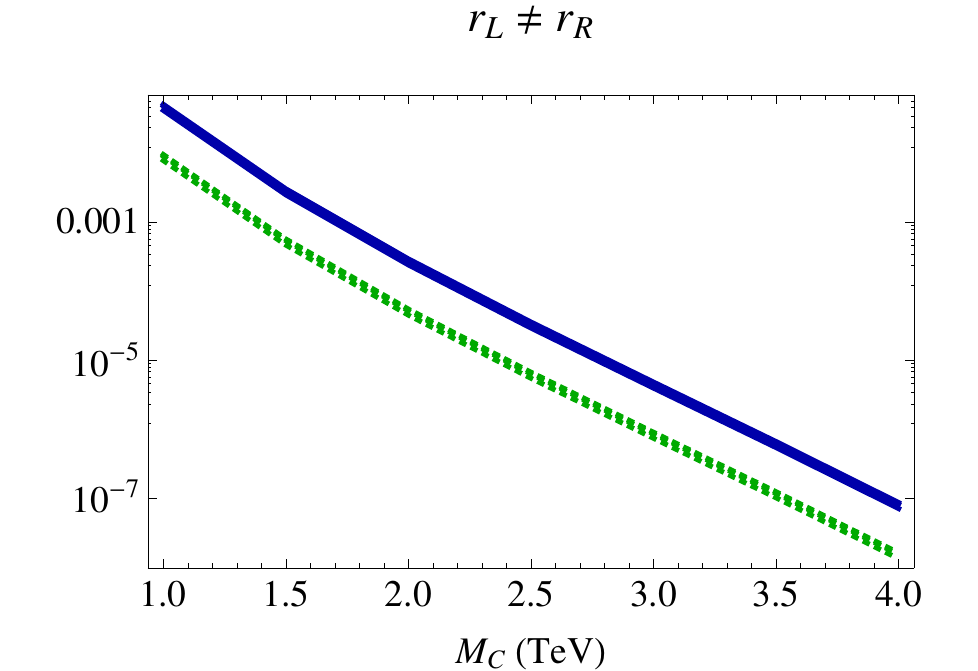}
\includegraphics[width=.329\textwidth]{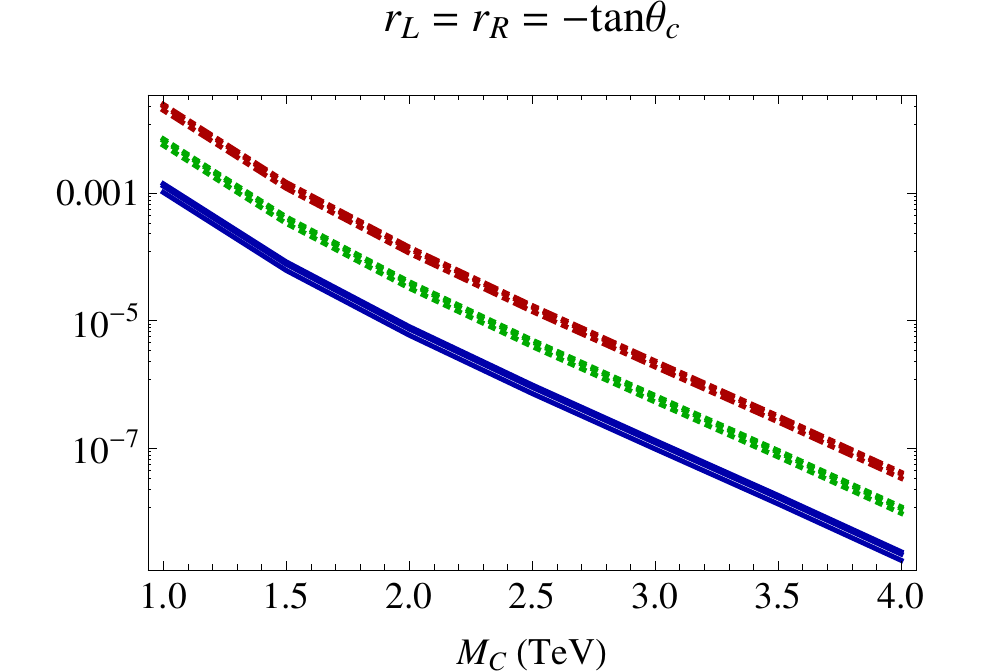}
\includegraphics[width=.329\textwidth]{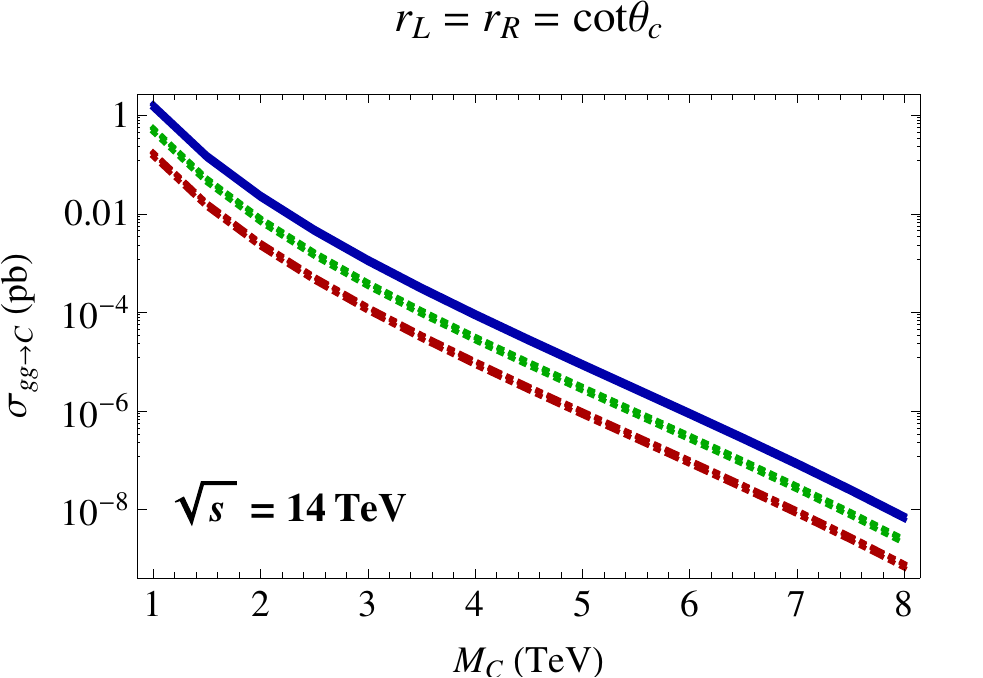}
\includegraphics[width=.329\textwidth]{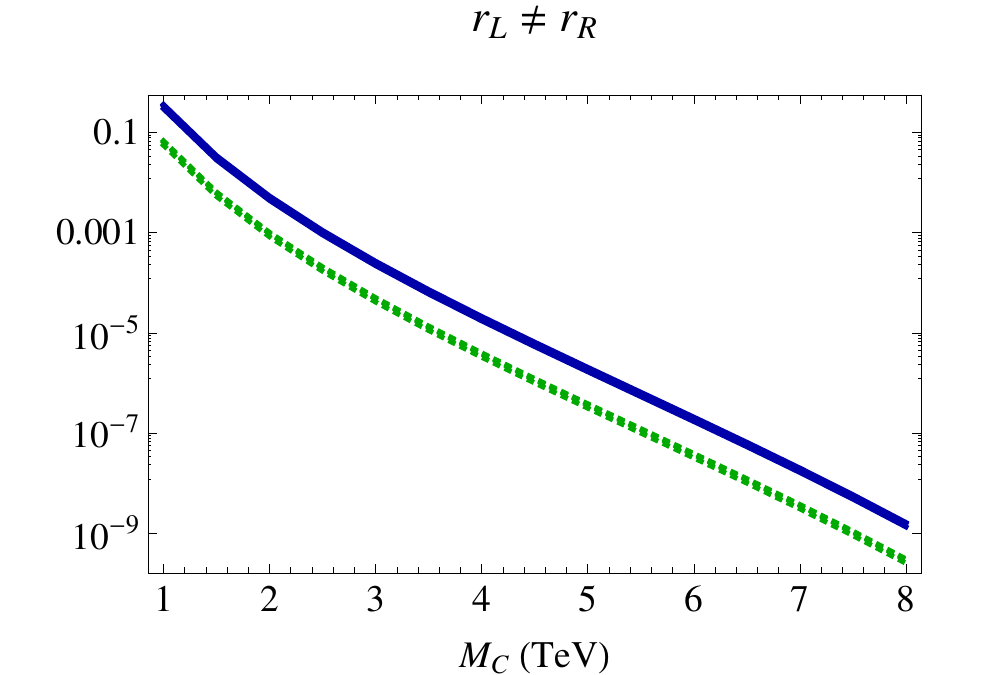}
\includegraphics[width=.329\textwidth]{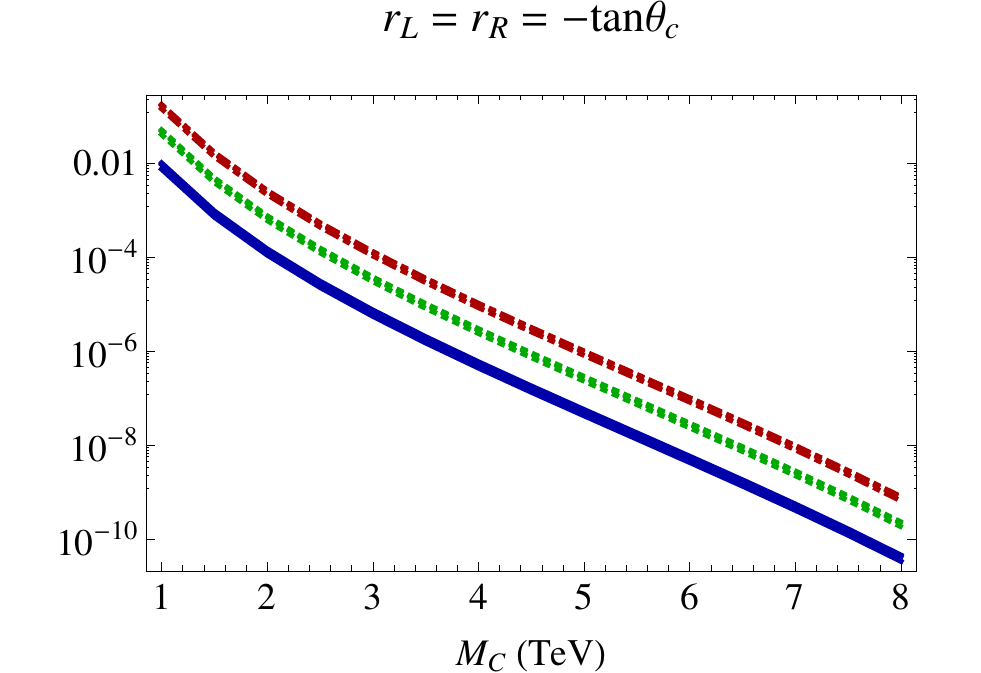}
\caption{Inclusive cross section foxr the gluon-gluon fusion to coloron process as a function of coloron mass at the LHC for CM energies of 8~TeV (top row) and 14~TeV (bottom row) with various (flavor universal) fermion charge assignments. Three representative values of the mixing angle have been plotted. The thickness of each curve reflects its (moderate) dependence on the $R$ parameter \eqref{R} with $1/2 \leq R \leq 2$, and the factorization scale, $\mu_F$, has been set equal to the coloron mass, $M_C$. Note the absence of a curve for the axigluon \cite{Frampton:1987dn,Bagger:1987fz}, corresponding to the maximal mixing $\sin^2\theta_c = 0.5$ and $r_L \neq r_R$, as the gluon fusion production of the coloron is forbidden by symmetry in this case and the production amplitude is identically zero (c.f. \eqref{iMfin}).}
\label{GGCCS}
\end{center}
\end{figure}

Fig.~\ref{GGCCS} illustrates the inclusive cross section for gluon-initiated coloron production \eqref{CSfin} as a function of the coloron mass at the LHC for two CM energies. The top row, corresponding to $\sqrt s = 8$~TeV, shows the cross section curves with various (flavor universal) fermion charge assignments (c.f. \eqref{H}) and three different values of the mixing angle, for a coloron mass range of $1 \leq M_C \leq 4$~TeV. As mentioned, this cross section also depends on the ratio of the spectator mass to the coloron mass via the $R$ parameter \eqref{R}. This dependence is, however, quite mild and is represented by the thickness of each curve for the chosen range $1/2 \leq R \leq 2$. The bottom row of the same figure displays the corresponding results for $\sqrt s = 14$~TeV and a coloron mass range of $1 \leq M_C \leq 8$~TeV. Note the enhancement in the production cross section, due to the higher gluon content of the protons at this CM energy.

It is worthwhile to note the absence of the maximal mixing curve ($\theta_c = \pi/4$) in the center plot that displays the case of unequal fermion charge assignment, $r_L \neq r_R$. This curve is associated with the axigluon \cite{Frampton:1987dn,Bagger:1987fz}, which posseses a parity symmetry exchanging $SU(3)_{1c}$ and $SU(3)_{2c}$, under which the gluon is even and the coloron is odd. The gluon fusion amplitude to produce a coloron, \eqref{iMfin}, therefore vanishes identically to all orders in perturbation theory. 

\begin{figure}
\begin{center}
\includegraphics[width=.329\textwidth]{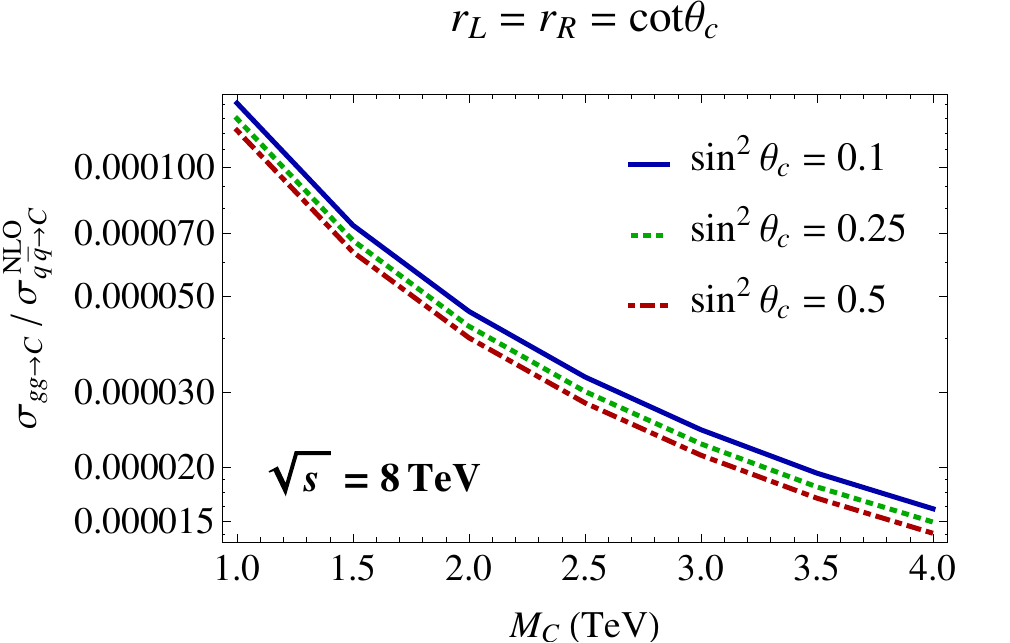}
\includegraphics[width=.329\textwidth]{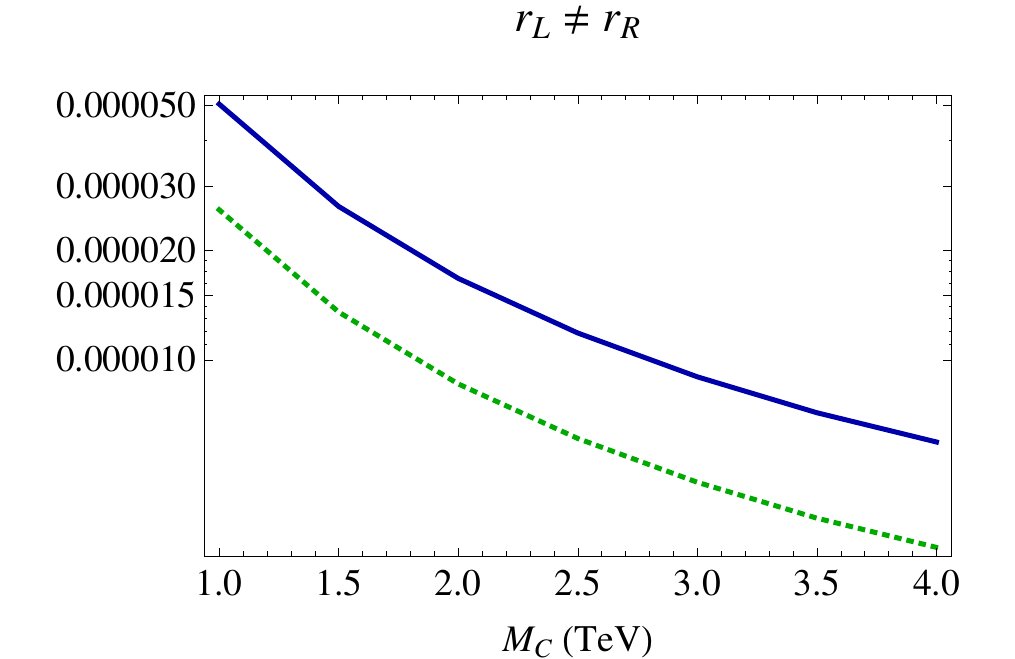}
\includegraphics[width=.329\textwidth]{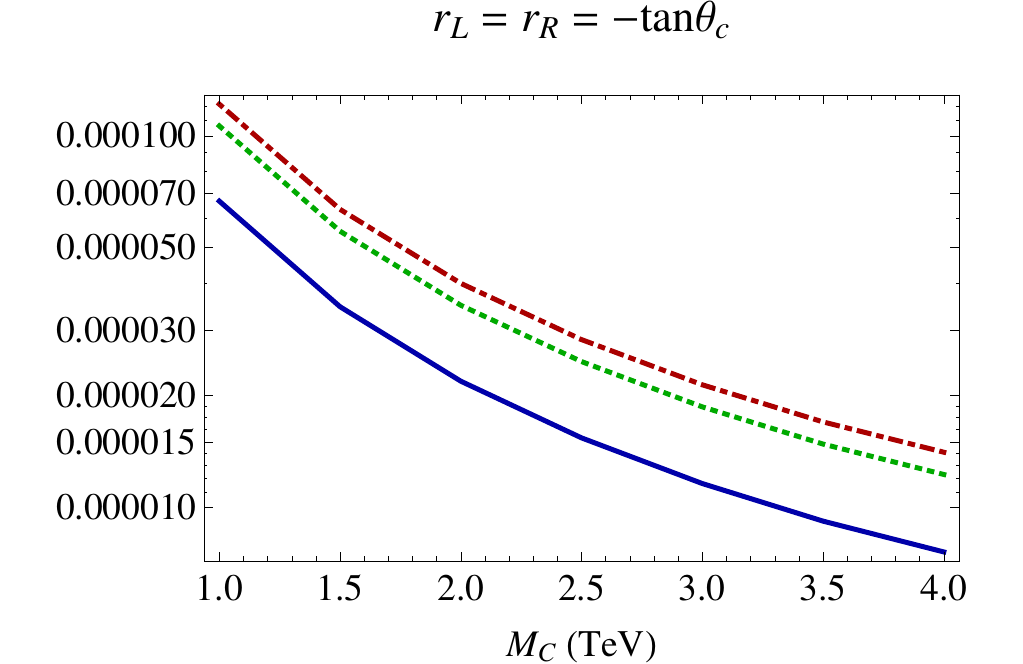}
\includegraphics[width=.329\textwidth]{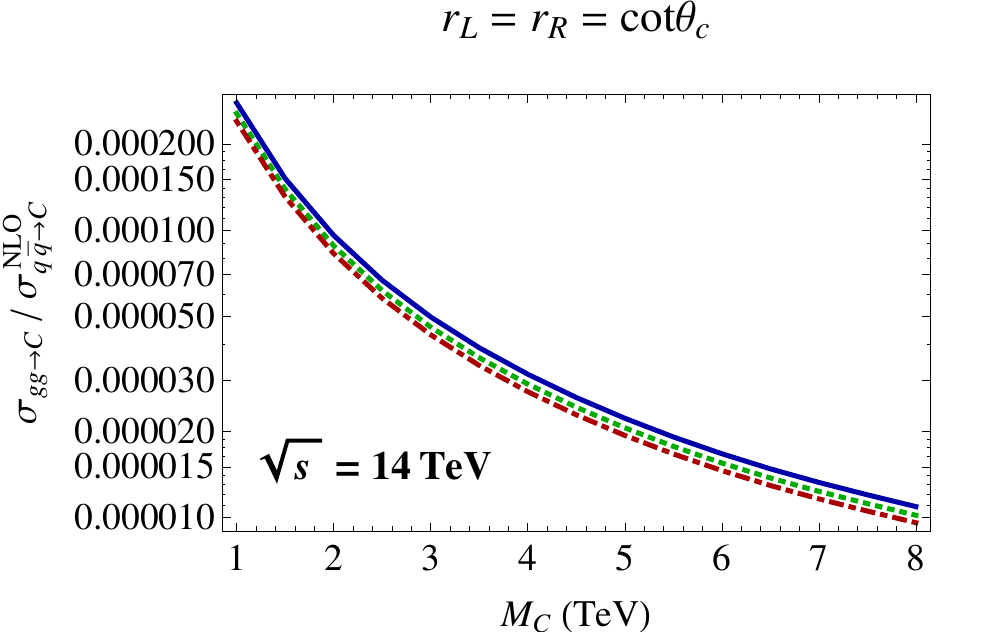}
\includegraphics[width=.329\textwidth]{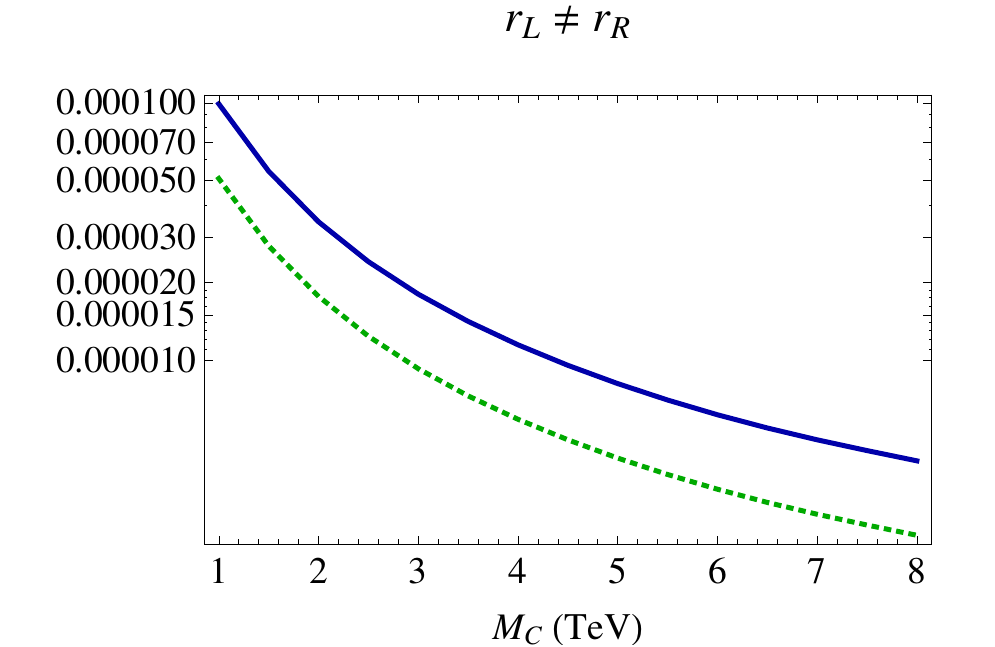}
\includegraphics[width=.329\textwidth]{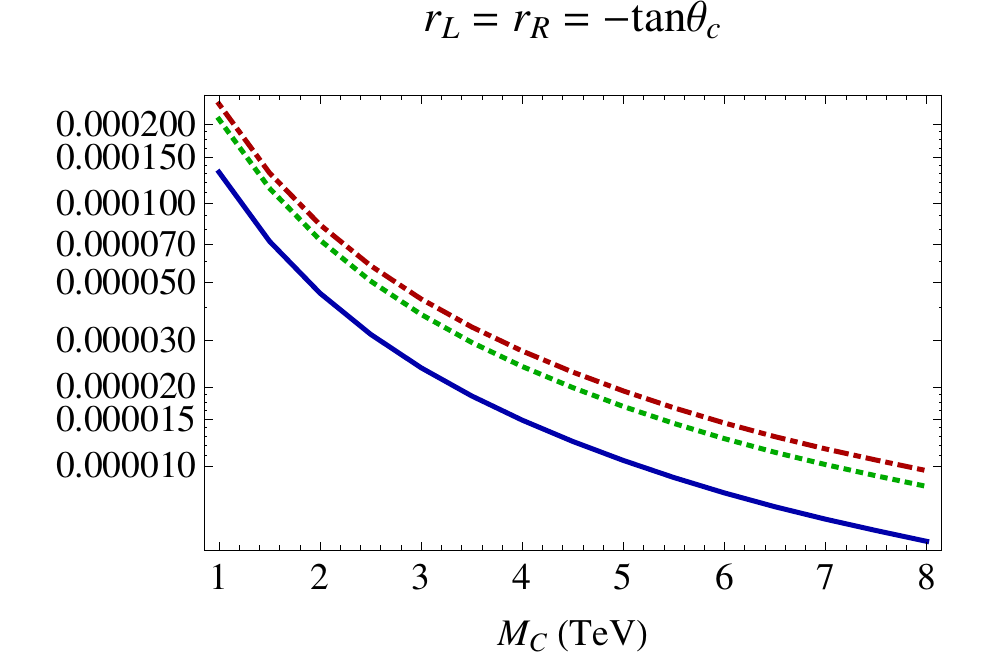}
\caption{Ratio of the coloron production cross section via gluon fusion with respect to the quark-initiated production channel, where the latter is evaluated at next-to-leading order \cite{Chivukula:2011ng}. Again, three representative values of the mixing angle have been plotted for the LHC CM energies of 8~TeV (top row) and 14~TeV (bottom row) with various fermion charge assignments. The factorization scale, $\mu_F$, has been set equal to the coloron mass as before, and we have chosen $R=1$ for illustration. Note, once more, the absence of the axigluon curve.}
\label{RatioNLO}
\end{center}
\end{figure}

In order to put the analysis of this production process into perspective, let us compare it with the dominant (tree-level) production channel, involving quarks. A comprehensive study of the latter at the next-to-leading order (NLO) is presented in \cite{Chivukula:2011ng}, where, in addition to the $q\bar{q} \to C$ one-loop virtual corrections, the (soft and collinear) real emission processes $q\bar{q} \to gC$, $qg \to qC$ and $\bar qg \to \bar qC$ are also taken into account. Fig.~\ref{RatioNLO} illustrates the ratio between the coloron production cross section via gluon-fusion and the NLO cross section for the quark-initiated channel at the LHC with two CM energies.\footnote{For brevity, in Fig.~\ref{RatioNLO} we have used the subscript $q\bar{q} \to C$ to indicate collectively all the mentioned NLO production processes containing one or more quarks in the initial state.} As before, three values of the mixing angle are displayed with various fermion charge assignments, and we have set $R=1$ for convenience. Once again, the curve for axigluon production is absent, due to the parity symmetry explained above. It is evident that the gluon-fusion production process is subdominant by four orders of magnitude by comparison with the quark-initiated production, even for the highest CM energy. 

We conclude that the gluon fusion contribution to coloron production is phenomenologically irrelevant except in the case of a fermiophobic coloron with a coupling of order ${\cal O}(10^{-2}g_s)$ or smaller.

\section{Updated NLO Coloron Production Rates at Hadron Colliders}
\label{Kfactors}

Finally, having shown that the gluon fusion contribution to coloron production is numerically insignificant (except in the case of fermiophobic colorons), we present an update of the NLO cross section for coloron production, as given in Eq.~(81) of \cite{Chivukula:2011ng}. As in that reference, we compute the `$K$-factor' for coloron production, defined as
\begin{equation}\label{K}
K \equiv \frac{\sigma^{\text{NLO}}_{q\bar q \to C}}{\sigma^{\text{LO}}_{q\bar q \to C}} \ ,
\end{equation}
where, $\sigma^{\text{NLO}}_{q\bar q \to C}$ is the NLO quark-initiated coloron production cross section and $\sigma^{\text{LO}}_{q\bar q \to C}$ is that for tree-level production, as given in \cite{Bagger:1987fz}.
The $K$-factors computed here differ slightly from those previously reported in \cite{Chivukula:2011ng}, largely due to the fact that here we use the modern CT10
\cite{Lai:2010vv} PDFs while the previously reported $K$-factors were calculated  using the Mathematica package for CTEQ5 \cite{Lai:1999wy}.\footnote{In addition, here we consistently apply the definition of $\theta_c$ in the coloron coupling in terms of $g_s$ extracted from $\alpha_s(M_Z)$, as described in Sec.\ref{subsec:hadron}.} 

\begin{figure}
\begin{center}
\includegraphics[width=.329\textwidth]{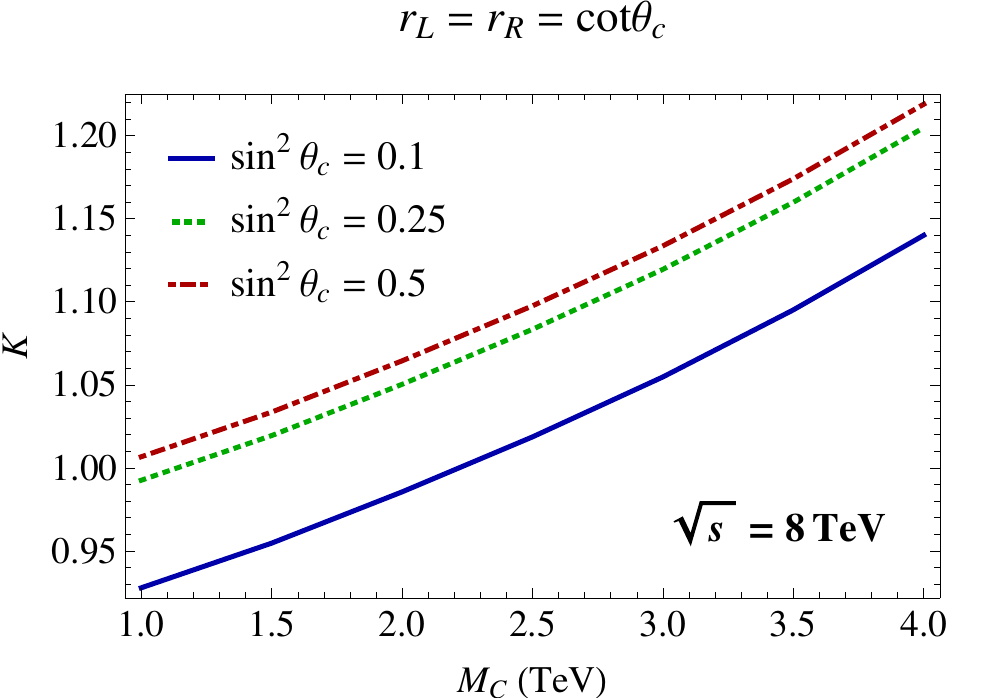}
\includegraphics[width=.329\textwidth]{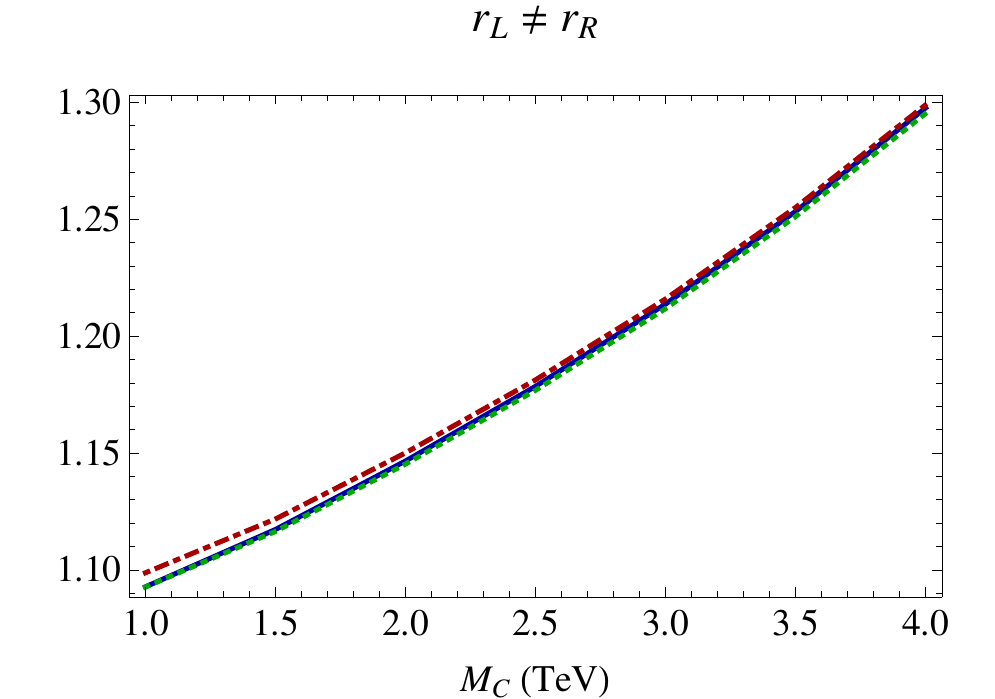}
\includegraphics[width=.329\textwidth]{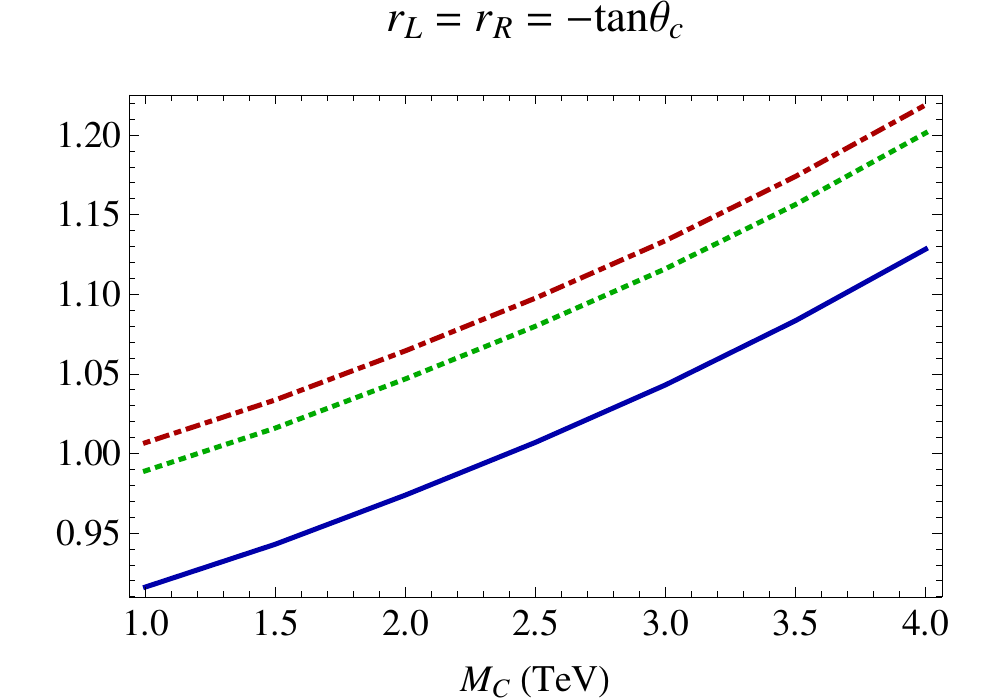}
\includegraphics[width=.329\textwidth]{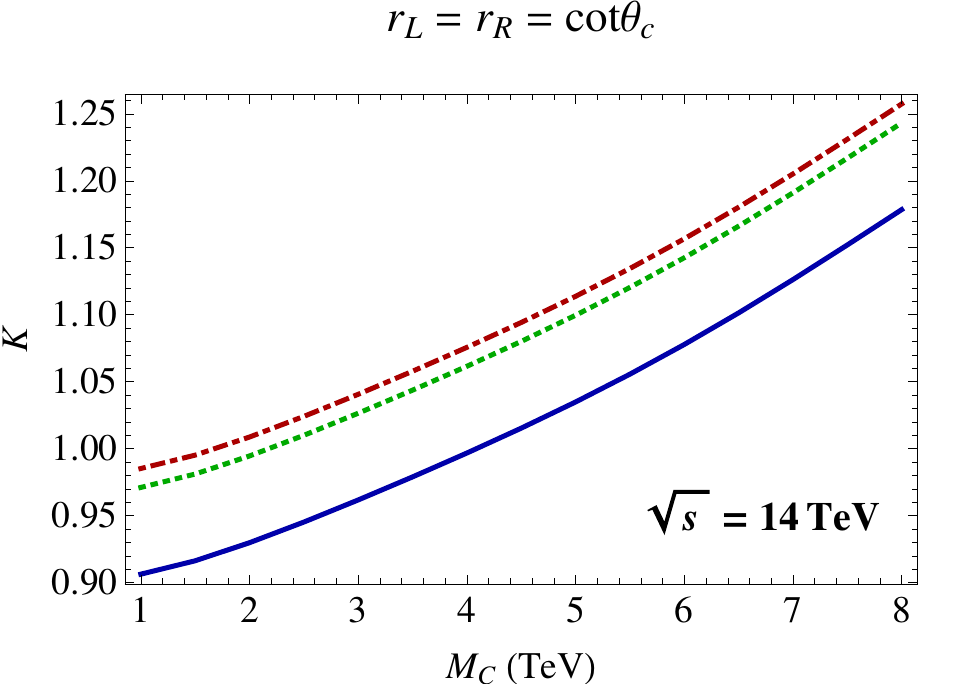}
\includegraphics[width=.329\textwidth]{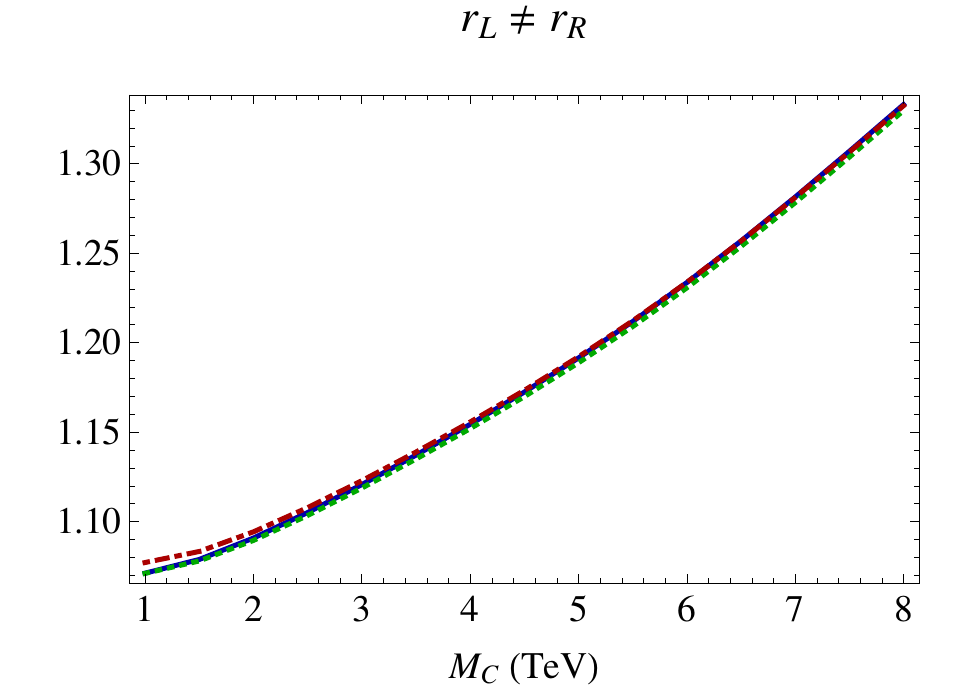}
\includegraphics[width=.329\textwidth]{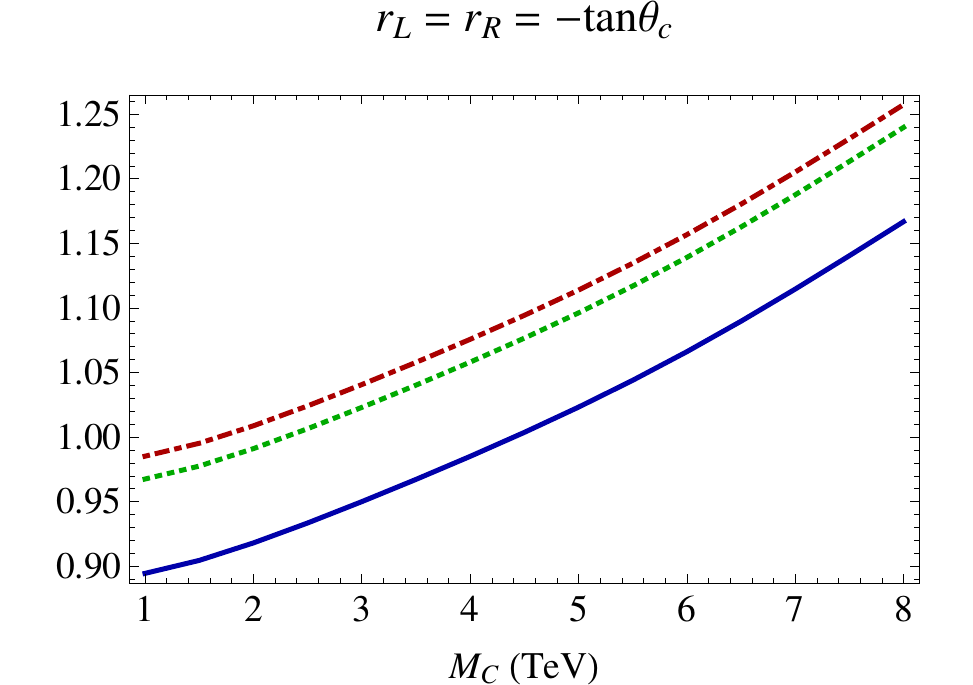}
\caption{The $K$-factor values \eqref{K} for two CM energies of the LHC, $\sqrt s = 8$~TeV (top row) and $\sqrt s = 14$~TeV (bottom row). Different fermion charge assignments with three values of the mixing angle have been plotted, for $\mu_F = M_C$. See also the tables of $K$-factors in Appendix~\ref{Kf}.}
\label{Kplots}
\end{center}
\end{figure}

The $K$-factor values \eqref{K} have been plotted in Fig.~\ref{Kplots} for the LHC with $\sqrt{s}=8$ and 14 TeV,\footnote{Plots of the $K$-factor for the LHC at 7 TeV may be found in \protect\cite{Chivukula:2011ng}.} with the three mixing angle values for different fermion charge assignments as before. We see that, as in \cite{Chivukula:2011ng}, the $K$-factor can be as large as 30\%. 
The numerical values of the $K$-factor associated with different choices of various parameters are tabulated in Appendix~\ref{Kf} for the LHC at beam energies of $\sqrt s = 7$, 8, and 14 TeV and also at the Tevatron for a beam energy of $\sqrt{s} = 1.96$~TeV.

\section{Conclusions}
\label{conclude}

In this paper we have presented results that complete the study of the next-to-leading order (NLO) QCD corrections to coloron production at the LHC and Tevatron begun in \cite{Chivukula:2011ng}. Our calculations apply directly to any model with an $SU(3)_{1c} \times SU(3)_{2c}$ gauge structure. They also apply approximately to the production of $KK$ gluons and colored technivector mesons to the extent that the $SU(3)_{1c} \times SU(3)_{2c}$
model is a good low-energy effective theory for models which incorporate these particles.
We used the pinch technique to investigate coloron production via gluon fusion. We demonstrated that this one-loop production amplitude is finite, and found that its numerical contribution to coloron production is typically four orders of magnitude smaller than the contribution from quark annihilation. Hence, the production of colorons via gluon-fusion is only relevant for (nearly) fermiophobic colorons. In addition, we have updated the results for the NLO QCD corrections to coloron production, and have presented plots and tables of our results for a range of coloron masses, mixing angles, and fermion charges at the Tevatron, the low-energy LHC and the high-energy LHC

\section*{Acknowledgments}

We would like to thank Roshan Foadi for useful discussions. We also thank Tomohiro Abe for his assistance with the Fortran code. This work is supported  in part, by the US National Science Foundation under grant PHY-0854889. J.R. was supported by the China Scholarship Council, and by the NSF of China (under grants 11275101, 10625522, 11135003). A.F. was supported by the Tsinghua Outstanding Postdoctoral Fellowship, and by the NSF of China (under grants 11275101, 11135003).

\appendix

\section{Feynman Rules\footnote{The Feynman rules presented here are derived in \protect\cite{Chivukula:2011ng}.}} \label{FR}

\begin{figure}
\begin{center}
\includegraphics[width=\textwidth]{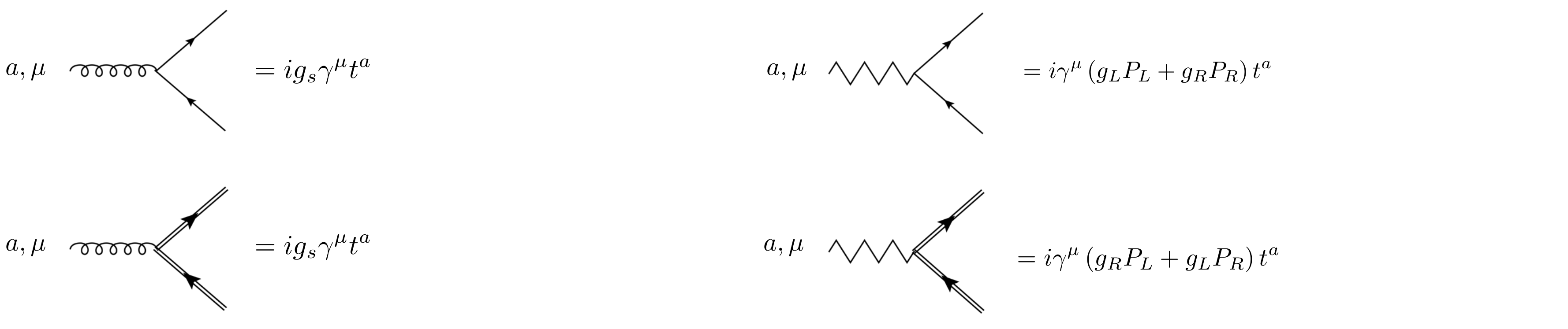}
\includegraphics[width=\textwidth]{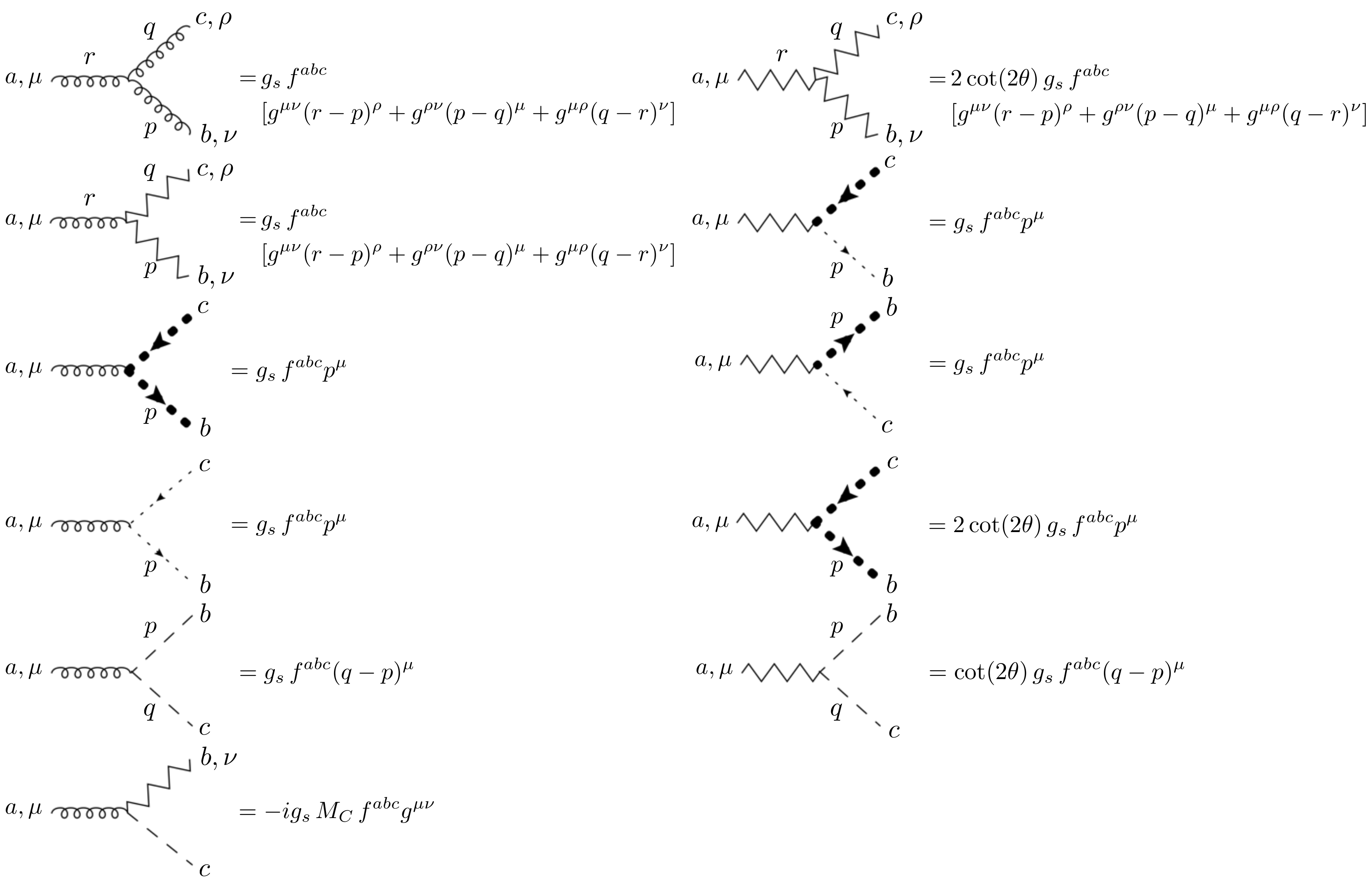}
\caption{Feynman rules for the trilinear vertices. In each diagram the momenta are toward the vertex.}
\label{fig:three}
\end{center}
\end{figure}

\begin{figure}
\begin{center}
\includegraphics[width=\textwidth]{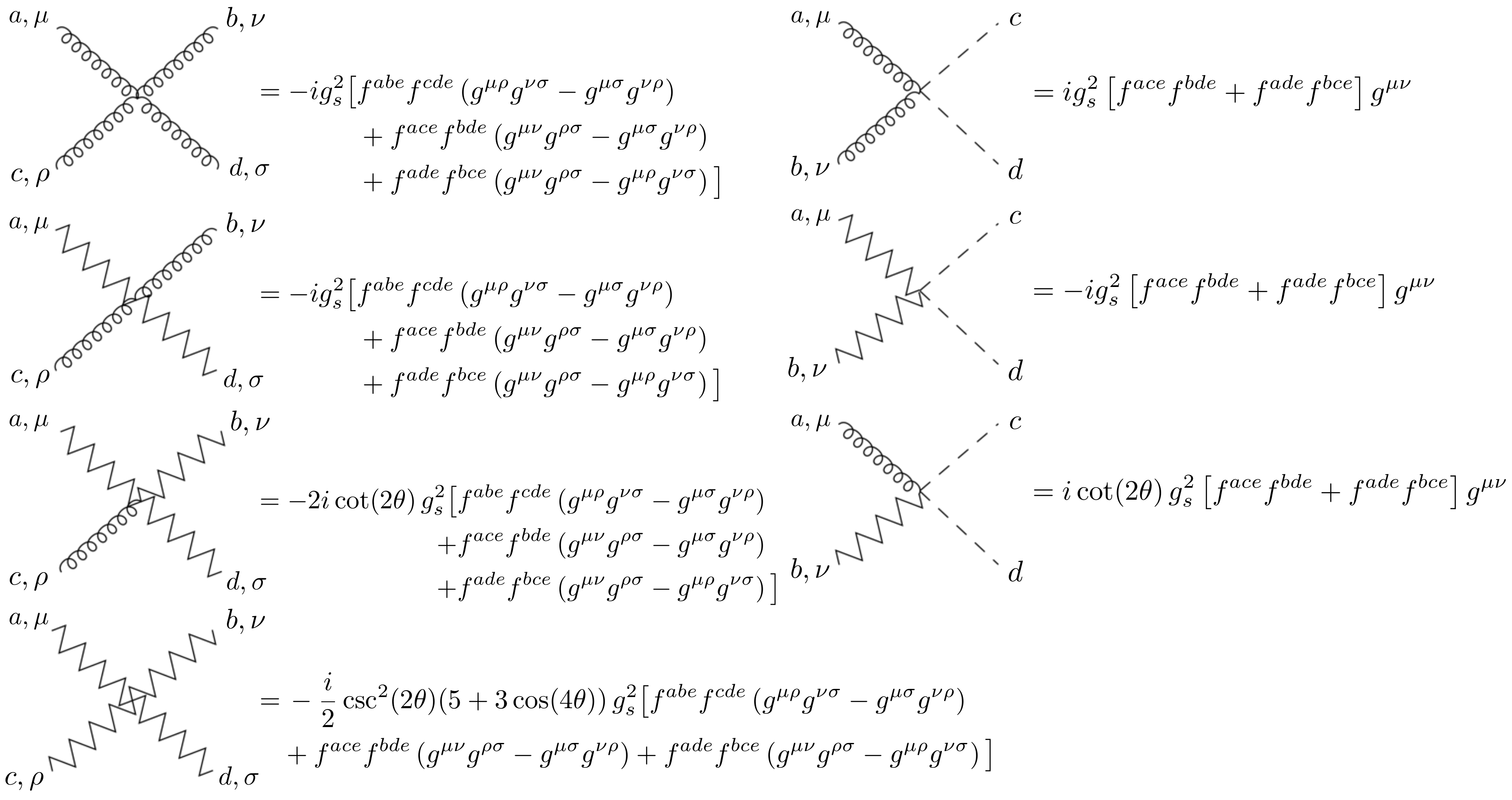}
\caption{Feynman rules for the quartic vertices.}
\label{fig:four}
\end{center}
\end{figure}

The Feynman rules for the trilinear and quartic vertices are shown in Figs.~\ref{fig:three} and~\ref{fig:four}, respectively. The coloron is represented by a zigzag line, the coloron ghost by a sequence of filled circles, and the eaten Nambu-Goldstone bosons by dashed lines. The spectator is depicted as a continuous double-line. All other particles are denoted as in QCD standard notation.

\section{Numerical Values of the $K$-Factor at the Tevatron and LHC} \label{Kf}

\begin{figure}
\begin{center}
\includegraphics[width=.329\textwidth]{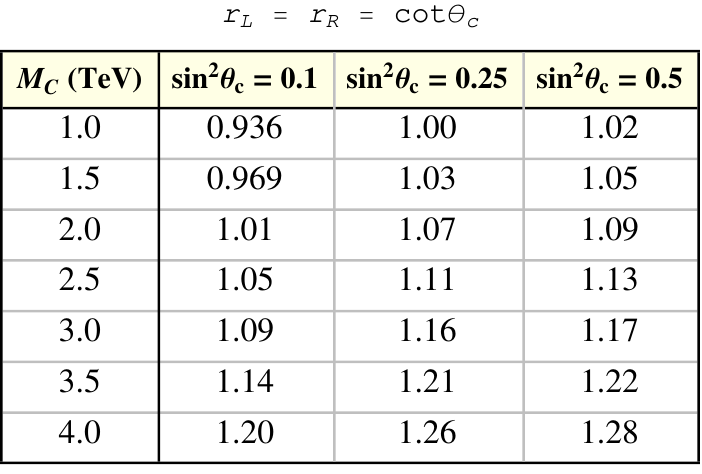}
\includegraphics[width=.329\textwidth]{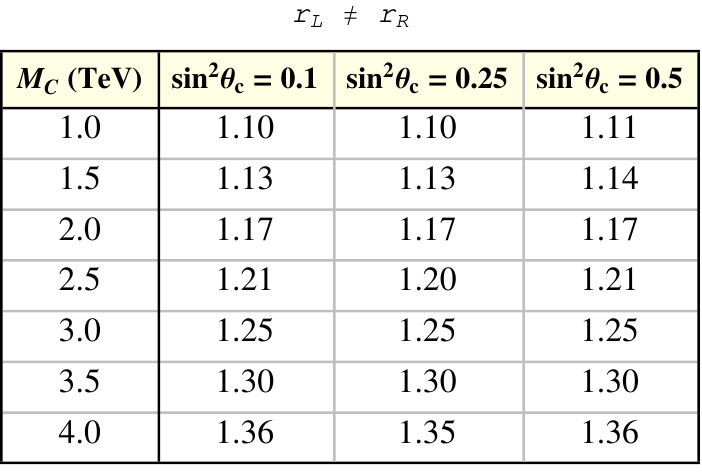}
\includegraphics[width=.329\textwidth]{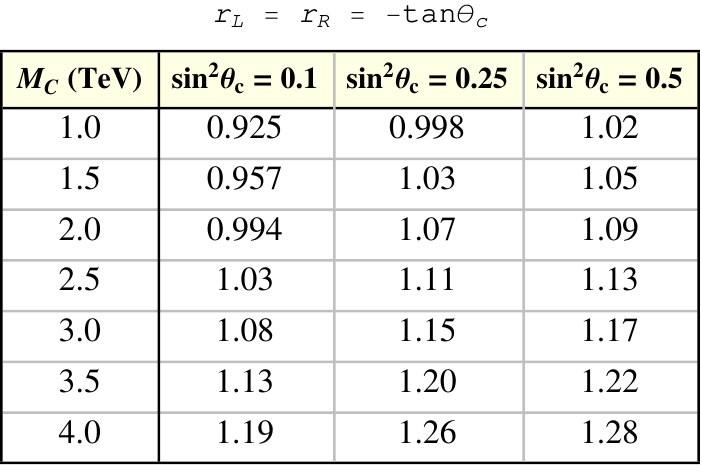}
\caption{LHC $K$-factor tables for colorons of various masses and mixing angles with $\sqrt s = 7$~TeV. Each table corresponds to a different charge assignment of the fermions. These values update those listed in \protect\cite{Chivukula:2011ng}, and the differences are largely due to our use here of updated CT10 structure functions \protect\cite{Lai:2010vv}.}
\label{KLHC7}
\end{center}
\end{figure}

\begin{figure}
\begin{center}
\includegraphics[width=.329\textwidth]{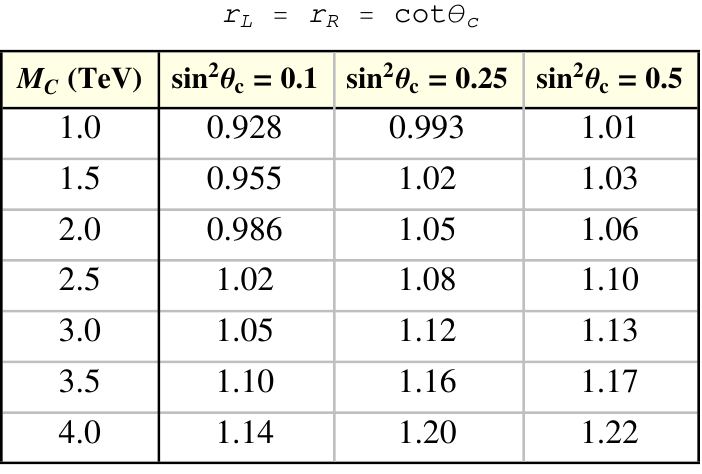}
\includegraphics[width=.329\textwidth]{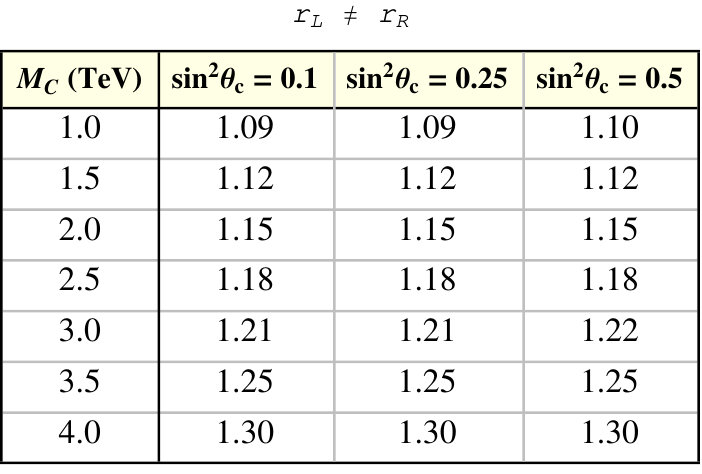}
\includegraphics[width=.329\textwidth]{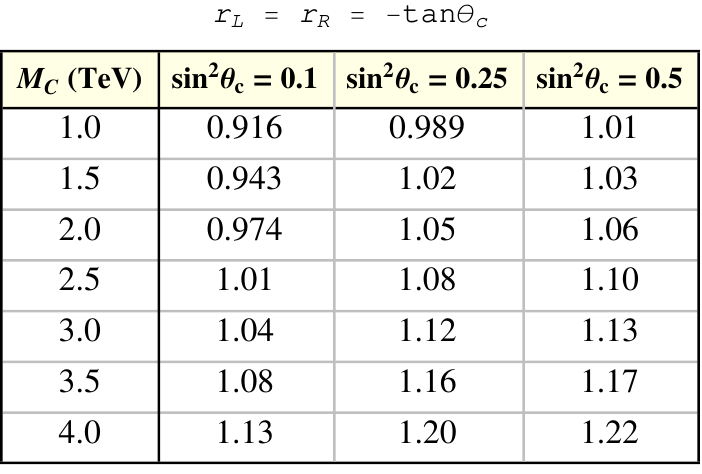}
\caption{LHC $K$-factor tables for colorons of various masses and mixing angles with $\sqrt s = 8$~TeV. Each table corresponds to a different charge assignment of the fermions.}
\label{KLHC8}
\end{center}
\end{figure}

\begin{figure}
\begin{center}
\includegraphics[width=.329\textwidth]{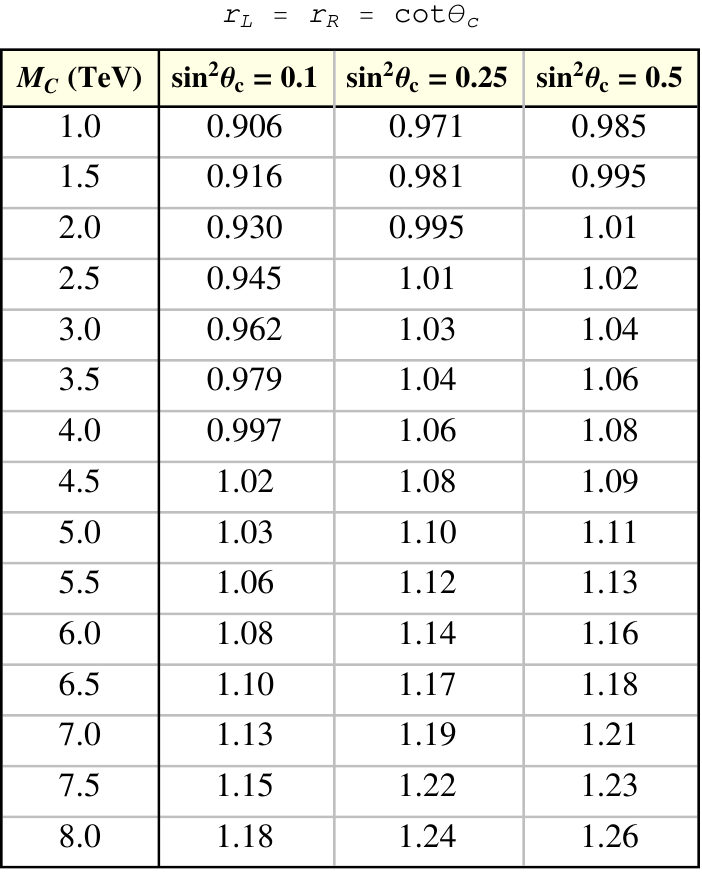}
\includegraphics[width=.329\textwidth]{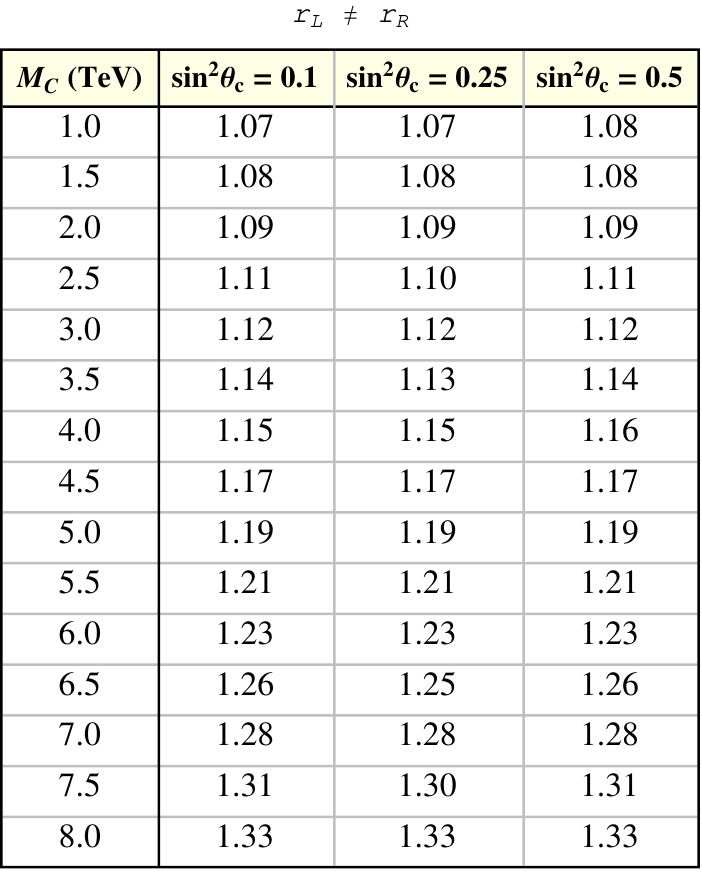}
\includegraphics[width=.329\textwidth]{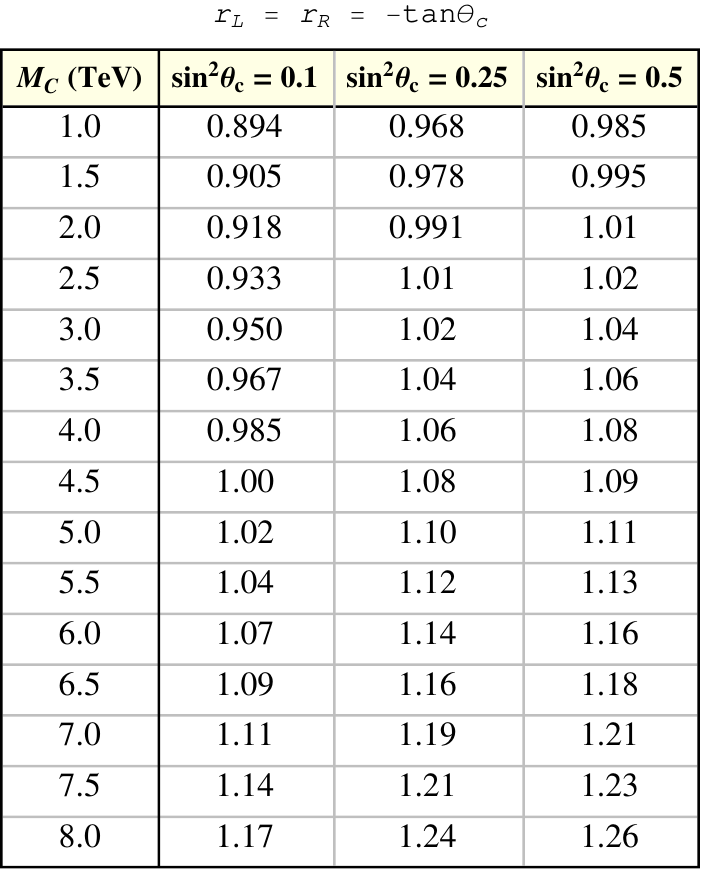}
\caption{LHC $K$-factor tables for colorons of various masses and mixing angles with $\sqrt s = 14$~TeV. Each table corresponds to a different charge assignment of the fermions.}
\label{KLHC14}
\end{center}
\end{figure}

\begin{figure}
\begin{center}
\includegraphics[width=.329\textwidth]{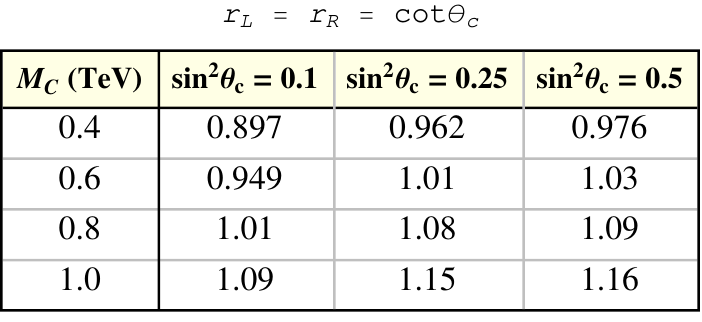}
\includegraphics[width=.329\textwidth]{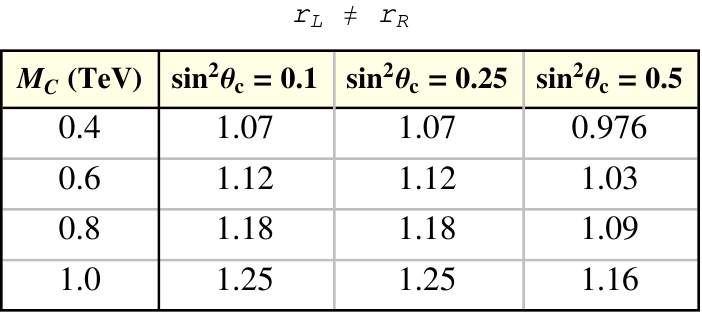}
\includegraphics[width=.329\textwidth]{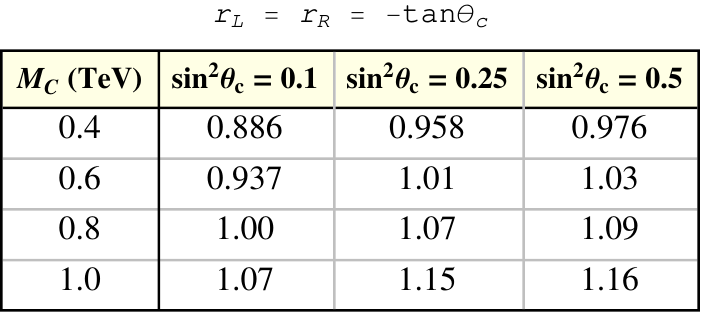}
\caption{Tevatron $K$-factor tables for colorons of various masses and mixing angles with $\sqrt s = 1.96$~TeV. Each table corresponds to a different charge assignment of the fermions.}
\label{KTEV}
\end{center}
\end{figure}

In this Appendix, we report the numerical values of the $K$-factor for the LHC and the Tevatron as a function of the coloron mass. The three LHC CM energies of $\sqrt s = 7$, 8 and 14 TeV have been presented in Figs.~\ref{KLHC7}, \ref{KLHC8} and \ref{KLHC14}, respectively, while the Tevatron $\sqrt s = 1.96$~TeV results are given in Fig.~\ref{KTEV}. The figures correspond to various fermion charge assignments, each containing three different mixing angles. 

\end{document}